\documentclass[pre,a4paper,twocolumn,showpacs,amsmath]{revtex4}
\usepackage{graphicx,amsfonts,amsmath,color,bm,epstopdf}

\newcommand{\ad}[1]{\tilde{#1}}
\newcommand{\tsx}{\tilde{\sigma}_{xx}}
\newcommand{\tsy}{\tilde{\sigma}_{yy}}

\newcommand{\tui}{\tilde{u}_i}

\newcommand{\exs}{e_{xx}^*}
\newcommand{\eys}{e_{yy}^*}

\newcommand{\eis}{e_{ij}^*}
\newcommand{\sis}{\sigma_{ij}^*}

\newcommand{\bsigma}{\bm{\sigma}}

\begin{document}
\title{Eshelby inclusions in granular matter: theory and simulations}
\author{Sean McNamara}
\author{J\'er\^ome Crassous}
\author{Axelle Amon}

\affiliation{Universit\'e de Rennes 1, Institut de Physique de Rennes
  (UMR UR1-CNRS 6251), B\^{a}t.~11A, Campus de Beaulieu, F-35042 Rennes, France}

\begin{abstract}
We present a numerical implementation of an active inclusion in a
granular material submitted to a biaxial test. We discuss the
dependence of the response to this perturbation on two parameters: the
intra-granular friction coefficient on one hand, the degree of the
loading on the other hand. We compare the numerical results to
theoretical predictions taking into account the change of volume of
the inclusion as well as the anisotropy of the elastic matrix.
\end{abstract}
\date{\today}
\maketitle

\section{Introduction}
Despite decades of intensive
research~\cite{schofield,nedderman,davis,vardoulakis}, the mechanics
leading to strain localization in soils are not well identified. For
geo-mechanicians, strain localization is the result of mechanical
instabilities~\cite{nedderman,davis,vardoulakis}. When the stress
exceeds a threshold (i.e. given by Mohr-Coulomb criterium for the
stress component ratio), the material fails. Before such failure,
soils flow plastically~\cite{schofield,davis,vardoulakis}, but the
link between this plastic flow and failure is quite
unclear~\cite{vardoulakis}. Recent approaches to describe the rheology
of granular materials propose non-local constitutive relations
inspired by models coming from soft glassy
materials~\cite{Kamrin2012,Bouzid2015} and introducing a new variable
to describe the local state of the material (the fluidity). The
physical interpretation of this fluidity is still unclear and
debated~\cite{Bouzid2015}

Metallic glasses are another example of materials where plastic flow
occurs before strain localization. It has been suggested that for
these amorphous materials~\cite{Argon1979a,Argon1979b}, the
macroscopic plasticity is the result of numerous elementary plastic
events. An elementary event may be viewed as a small and local
reorganization at the molecular scale.  Such events
have been found in numerical simulations
~\cite{Falk1998,Maloney2006,Tanguy2006}. This description in term of
elementary plastic events seems also relevant for soft glassy
materials (such as dense emulsions or foams) or compressed granular
materials~\cite{Maloney2006}. Such elementary events have been indeed
reported in some experimental
studies~\cite{Schall2007,Amon2012,Jensen2014,desmond2015,sentjabrskaja2015}. 

A very important consequence of those elementary plastic events is
that they are able to interact mechanically. Indeed, when they occur, they redistribute stress
  in the material~\cite{Picard2004}. For an elastic material, this
redistribution is long-ranged and anisotropic, so that it may trigger
other rearrangements further in the material along preferential
directions. This coupling between plastic events is believed to be at
the origin of formation of spatial
heterogeneities~\cite{nicolas2014,puosi2015,lin2014,lin2015} (such as
shear band) at the macroscopic scale.

Consequently a good understanding of the stress redistribution around
a single event is needed. It is a subject that has been already deeply
studied, especially
numerically~\cite{Kabla2003,Maloney2006,Tanguy2006}, while
experimental proofs of this redistribution are
scarce~\cite{Schall2007,LeBouil2014b,desmond2015}. Those studies show
that the stress redistribution around a local event may be well
described using a computation done by Eshelby~\cite{Eshelby1957} where
an inclusion embedded in an elastic matrix spontaneously changes
shape.  The analytical solutions obtained from such a calculation
provide a good description of the stress field measured numerically
around plastic events, i.e. a quadrupolar stress redistribution. Using
this theoretical representation of plastic events as Eshelby
inclusions and considering the coupling between several events,
predictions of the yield strain can be
obtained~\cite{Dasgupta2013a,Dasgupta2013b}, as well as of the
inclination of shear bands~\cite{Ashwin2013,LeBouil2014b}. Active
rearrangements have also been studied numerically in order to
investigate the dynamics of the response of the material to a local
rearrangement~\cite{Puosi2014} as well as the correlation between this
elastic response and the subsequent plastic
events~\cite{Priezjev2015}.

It has to be noted that most of the numerical studies discussed above
have been done in Lennard-Jones glasses for which a domain of truly
elastic response exists. So, an elastic response of the matrix
surrounding a small plastic event is not a surprise. For materials
which do not behave as perfectly elastically, the pertinency of
Eshelby's calculation can be questioned. Wu~{\it et al.} show that the
stress relaxation in liquid at times shorter than the Maxwell
relaxation time may be described accordingly to Eshelby's stress
tensors~\cite{wu2015}. The fourfold pattern characteristic of
Eshelby's stress tensors has observed in recent numerical studies of
granular flows at constant volume~\cite{guo2014}. Also, fluctuations
of strains are observed experimentally~\cite{LeBouil2014b} with
inclinations related to the maximum of the deviatoric part of
Eshelby's stress tensors.  Similar oriented patterns were observed in
numerical studies of compressed granular material~\cite{Kuhn1999},
although the author does not link them to the Eshelby's tensors.

The aim of this study is to show that the stress redistribution due to
a local reorganization in a granular material may be unambiguously
described using the Eshelby formalism. For this, we focus on a
realistic situation, where the granular material evolves at fixed
pressure, allowing compaction or dilatation. We implement in a
discrete element numerical model of a bidimensional frictional
granular assembly an active inclusion with a new procedure very close
to the spirit of the demonstration of Eshelby. We show that the
observed response of the granular assembly corresponds to the one
expected for an elastic material~\cite{Eshelby1957}. We study specifically the role of two parameters on the
response of the system: the value of the friction coefficient, which
governs the prevalency of sliding contacts in the system, and the
loading state of the sample, i.e. its proximity to failure.

This study is organized in the following way:

In section~\ref{sec:prep} we explain the mode of loading of the
material, which consists in a biaxial test, and we describe the
preparation of the numerical experiments. We also describe the typical
response obtained during the loading of the material. In
section~\ref{sec:theory}, we give an overview of the principle of the
calculation of Eshelby and we discuss our method of generating an
Eshelby-type inclusion in our simulations. In
section~\ref{sec:results}, we describe the stress field observed
around an active inclusion at different stages of the loading and for
different values of the friction coefficient. In
section~\ref{sec:discussion} we measure the angular distribution of
the anisotropic response and we present the results of the evolution
of the inclination of the maximum of the stress distribution as a
function of the volumetric strain. We discuss the modification of the
response of the system when the amount of plasticity in the granular
matrix increases, either because of decreasing friction or because the
system is closer to failure.

\section{Geometry and preparation of the numerical experiment}
\label{sec:prep}

In this section, we present the discrete element numerical simulations
and explain the method of preparation of the samples. We consider a
bidimensional sample submitted to a biaxial test. This classical
configuration~\cite{SoilMech} consists in submitting a granular sample
to well-controlled stress conditions without control of the
volume. Experimentally, a uniaxial compression is exerted along one
direction while a confining pressure is imposed on the lateral side of
the granular sample. Those kind of tests are classically used in soil
mechanics to study the strain-stress properties of granular
materials~\cite{SoilMech}. For large enough initial volume fraction,
localization of the deformation is observed above a
threshold~\cite{Desrues2015}, with the formation of shear
bands. Concerning the volumetric strain, still for initial volume
fraction larger than a critical value, some compaction is observed at
the beginning of the loading, followed by dilatancy at larger
strain. Those kinds of test are unusual for physicists who prefer to
work at constant volume and uncontrolled pressure, in order to study
the behavior of a sample at a given volume
fraction~\cite{Majmudar2007}. Yet, confining pressure controlled
set-ups are the most natural configurations when trying to understand
real life granular mechanical behavior.  Because we are interested in
this kind of configuration, we will give special care to local change
of volume in the following. As a matter of fact dilatancy and plastic
deformation are closely linked in granular materials.

In Sec.~\ref{sec:setup} and~\ref{sec:procedure}, we discuss the
numerical setup and test procedure.  In Sec.~\ref{sec:tests} we
describe general properties of our biaxial tests.

\subsection{Numerical Setup}
\label{sec:setup}

\begin{figure}
\centering
\includegraphics[width=0.8\columnwidth]{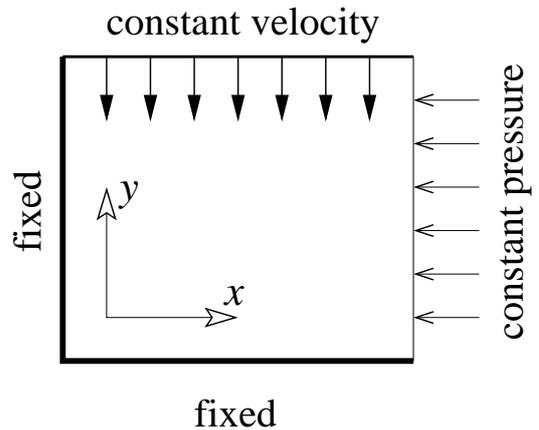}
\caption{Set up for the numerical simulations.  $N$ grains are
  contained within four walls: the bottom and left walls are fixed,
  the upper wall moves downward with constant velocity.  The right
  wall is mobile, and subjected to a constant pressure $p_0$.
  Throughout the paper, the coordinate axes $x$ and $y$ are oriented
  as shown.}
\label{fig:setup}
\end{figure}

We perform numerical simulations of a two-dimensional biaxial test.
The boundary conditions are sketched in Fig.~\ref{fig:setup}.  $N$
grains of total mass $M_0$ are confined by four perpendicular walls.
The left and bottom walls are fixed and motionless.  The top wall
moves downward with a constant velocity.  These three walls can be
considered to be of infinite mass, for their positions are not
modified by the forces exerted by the grains.  The right wall,
however, is mobile, with a mass of $M_0/100$.  It is kept in place by
a constant pressure $p_0$.

The grains are polydisperse disks
($r_\mathrm{max}/r_\mathrm{min}=1.5$) of uniform (two-dimensional)
mass density $\rho_0$.  The quantities $M_0$, $\rho_0$, and $p_0$
establish the system of units.  Thus length is measured in units of
$L_0=\sqrt{M_0/\rho_0}$, time in units of $T_0=\sqrt{M_0/p_0}$, and
velocity in units of $V_0=\sqrt{p_0/\rho_0}$.

The grains interact via traditional linear, damped springs in the
normal and tangential directions (normal spring constant
$k_N=2000p_0$, tangential spring constant $k_T=k_N/2$).  Weak linear
damping is also included in these interactions.  The grains also have
a (microscopic) friction ratio $\mu_m$: the tangential force may not
exceed $\mu_m$ times the normal force.  A weak rolling resistance is
added so that all vibration modes will be damped.  Finally, a weak
gravitational force is added to gently push ``rattlers'' against the
granular skeleton.

All four walls are frictionless, exerting only normal forces.  The
mobile wall has the same stiffness as the grains, but the three walls
of infinite mass are soft (spring constant $k_\mathrm{wall} =
k_N/\sqrt{N}$) and dissipative.  This novel boundary conditions
efficiently removes vibrational energy from the system without
resorting to global damping \cite{Particles2013}.

\subsection{Test procedure}
\label{sec:procedure}

The simulations presented below are done with $N=256 \times 256 =
65536$ grains.  The initial state is formed by compressing a granular
gas without friction (i.e. we set $\mu_m=0$), yielding a relatively
high density.  Then, at the beginning of the test, friction is turned
on (i.e. we set $\mu_m$ to its final value), and the velocity of the
top wall is set to $2 \times 10^{-5} H_0/T_0$, where $H_0 \approx 1.1
L_0$ is the initial height of the system. We checked that decreasing
further the velocity do not change the results of the simulation. As
the simulation proceeds, the entire state of the system (particle
positions and velocities, contact forces) are recorded at strain
increments of $\Delta\varepsilon_{yy} = 10^{-5}$, where the
deformation $\varepsilon_{yy}$ is defined in \ref{sec:tests}

\subsection{Global properties}
\label{sec:tests}

\begin{figure}[tbp]
\includegraphics[width=\columnwidth]{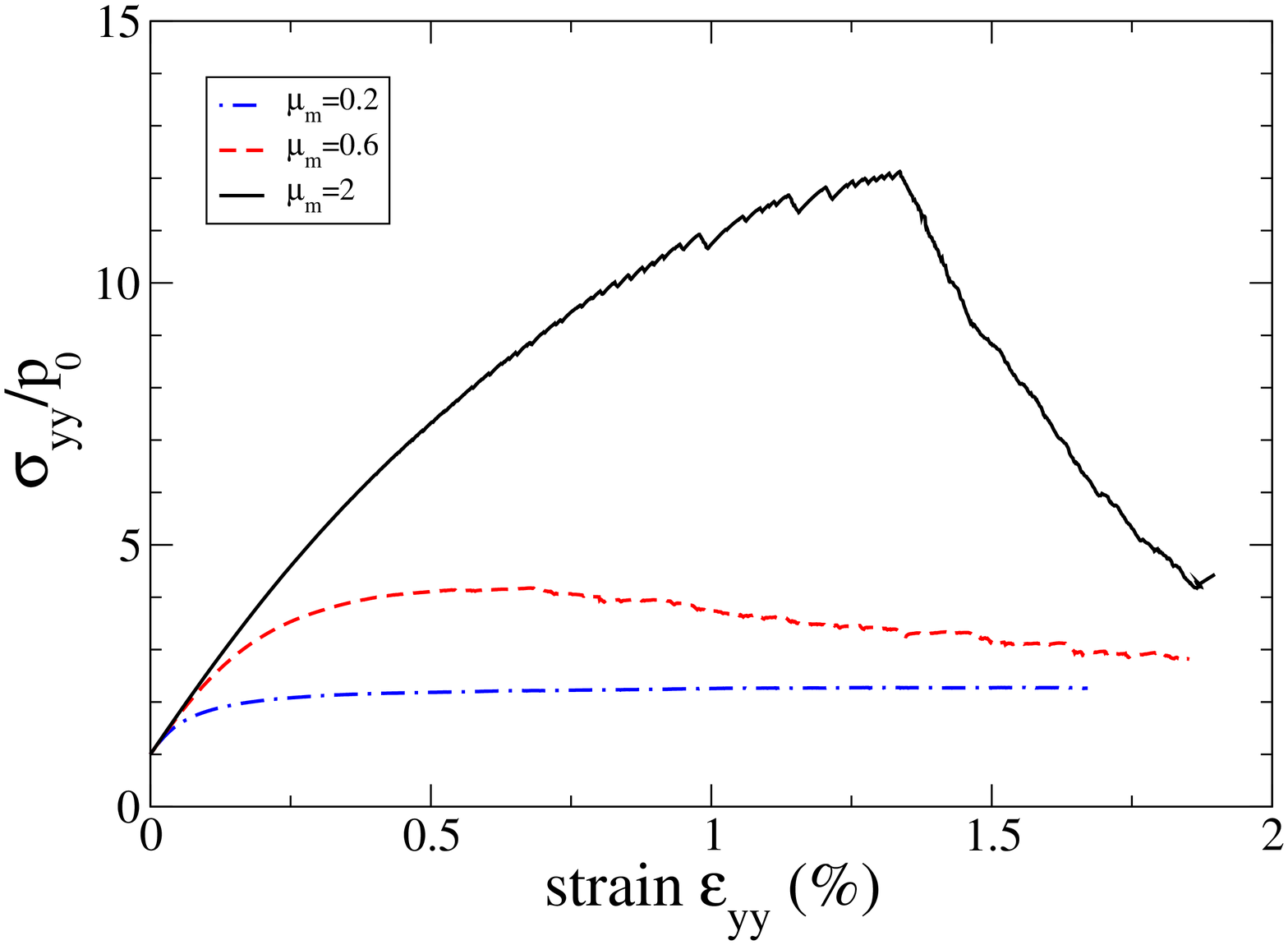}
\caption{Stress strain curves for the simulations analyzed in this paper.
}
\label{fig:stress-strain}
\end{figure}

In Fig.~\ref{fig:stress-strain}, we show the stress-strain curve for
the simulations studied in this paper.  The stress is deduced from the
force exerted by the grains on the upper (constant velocity) wall.
The strain is usually deduced in simulations from the position of this
wall, but that does not work in this situation due to the softness of
the wall.  Indeed, more than half of the apparent deformation can
occur at the wall, and not within the packing.  Instead, we
return to the definition of strain: $\varepsilon_{yy} = \partial
u_y/\partial y$, where $u_y$ is the $y$-component of the displacement
of a material point relative to a reference state.  We therefore
calculate the displacement of each grain, and accumulate $N$ pairs
$(y,u_y)$, and then do a linear regression and take the slope of as
$\Delta\varepsilon_{yy}$, the deformation from the reference state.
At the beginning of the simulation, the initial
condition is the reference state, but each time the deformation increases by
$10^{-5}$, the reference state is updated.
The total deformation, averaged over the full sample, is $\varepsilon_{yy}$
in Fig.~\ref{fig:stress-strain} and in the rest of the paper.

Fig.~\ref{fig:stress-strain} shows that we can control the peak stress
by changing the friction ratio $\mu_m$.
For $\mu_m=0.2$ and $\mu_m=0.6$,
we obtain results similar to those obtained experimentally
or in other numerical experiments.
When $\mu_m=2$, the peak stress
exceeds $10$ times the confining pressure.  This quite high value do not correspond to experimental values of grain-grain friction coefficients. However, it is an interesting limit case for theoretical reasons.

\begin{figure}[tbp]
\includegraphics[width=\columnwidth]{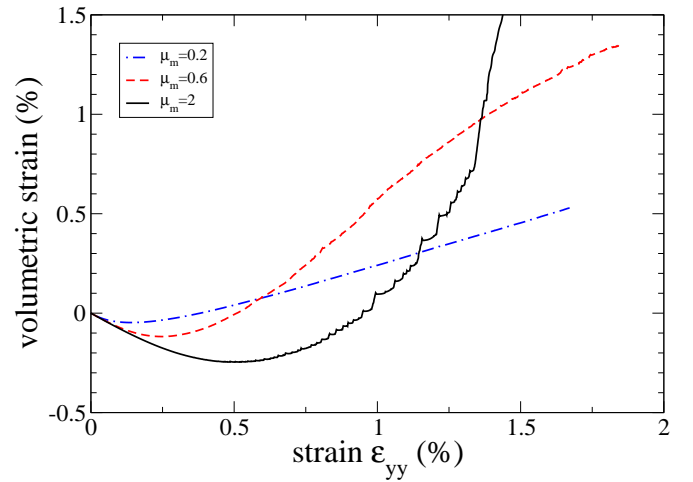}
\caption{Volumetric strain $\varepsilon_{xx}+\varepsilon_{yy}$
for the tests studied in this paper.}
\label{fig:volume-strain}
\end{figure}

In Fig.~\ref{fig:volume-strain}, we show the volumetric strain
$\varepsilon_{xx}+\varepsilon_{yy}$ during the tests.
(The strain component $\varepsilon_{xx}$ is
determined in the same way as $\varepsilon_{yy}$ strain component.)
This result is in agreement with the
experimental observations in biaxial and triaxial tests: the loading
curve presents a maximum and the volumetric strain first decreases
before continuously increasing.

\section{Eshelby's inclusion: theory and numerical implementation}
\label{sec:theory}

In this part we present our method for implementing an active
inclusion in a granular material. Active inclusions in numerical
simulation have been considered recently~\cite{Puosi2014,Priezjev2015}
in two-dimensional Lennard-Jones glasses by imposing displacements on
the particles forming the inclusion.  This approach amounts to
imposing a strain on the inclusion.  Here we propose a new procedure
to implement an active inclusion where a stress is imposed.  We do
this for two reasons.  First of all, we are dealing with granular
materials, where changes of volume are very important.  Thus we want
to leave the inclusion free to choose the volume it wants.  Second,
the stress we impose appears at the beginning of the Eshelby's
calculation~\cite{Eshelby1957}.  We will explain this more precisely
below.

Eshelby gave a general
method based on the linear superposition of small deformation
elasticity to solve a class of problems implicating an inclusion in an
elastic matrix. He also gives analytical solutions of several typical
problems. In the following, we will use the solution of
this problem: consider a small part of an infinite elastic
matrix which we will call the inclusion. Suppose that this inclusion
has a spontaneous change of shape. The Eshelby solution of interest
for us gives the stress field distribution in the matrix due to that
local event when the unconstrained spontaneous change of shape of the
inclusion is known.

The coordinate system used in the following is given in
Fig.~\ref{fig:setup}. For the vector positions, we use the polar
coodinates $\vec{r} = r \vec{n} = r \left( \begin{array}{c} \cos
  \theta\\ \sin \theta \end{array} \right)$. The origin is taken at the
center of the inclusion. In all the considered cases, the imposed
stress at the boundaries corresponds to biaxial stress-imposed
configuration:
\begin{equation*}
\mathbf{\sigma^{\infty}} = \left(
\begin{array}{cc}
\sigma^{\infty}_{xx} & 0\\
0 & \sigma^{\infty}_{yy}\\
\end{array}
\right)
\end{equation*}
with $\sigma^{\infty}_{xx} \text{ and } \sigma^{\infty}_{yy}<0$, and
$\sigma^{\infty}_{xx} - \sigma^{\infty}_{yy} >0$.\\

In this theoretical section, we first recall the general principle of
Eshelby's calculation. In the second part, we give the
general form of the analytical solution far from the inclusion. In the
last part our method of generating Eshelby-type inclusion in the
numerical simulation is described.

\subsection{Principle of the calculation}
\label{sec:principle}
Consider an inclusion, within an infinite elastic matrix, which
undergoes a spontaneous change of shape. If it had not been inside a
matrix, this change of shape would correspond to the homogeneous
strain tensor $\eis$. The elastic stress field associated to this
strain tensor is $\sis = \lambda e_{kk}^* \delta_{ij}+ 2\mu \eis$,
$\lambda$ and $\mu$ being the Lam\'e coefficients~\cite{Landau}.

As the inclusion is in a matrix, its change of shape will interact
with the elastic surrounding medium, leading to a displacement field
$\tui (\vec{r})$ in the matrix, generated by the rearrangement and
given in 2D by:
\begin{equation}
\tui (\vec{r}) = \int_\mathcal{C}  G_{ij}^{2D}(\vec{r} - \vec{r'})
\sigma_{jk}^* dc_k'
\label{eq:uGreen2D}
\end{equation}
where the integral covers the inclusion contour $\mathcal{C}$,
$G_{ij}^{2D}(\vec{r} - \vec{r'})$ is the bidimensional Green function
giving the displacement response awaited at $\vec{r}$ from a ponctual
unitary force exerted in $\vec{r'}$ belonging to the contour on an
element $dc_k'$~\cite{Landau,Eshelby1957}.

The following part discusses the far-field redistributed deviatoric
stress for $\eis$ coaxial with the loading
$\sigma^{\infty}_{ij}$. Complete solutions (with near-field terms) of
the previous equations in the case of an elliptical inclusion can be
found in~\cite{Ashwin2013}.

\subsection{Coaxial solutions} \label{subsec:2D}
The displacement field in the matrix far from the inclusion can be
approximated by (see Appendix~\ref{sec:Ap_2D}):
\begin{equation}
\tilde{u_i}(\vec{r}) = \frac{S}{2 \pi r} \frac{\lambda +
  \mu}{\lambda + 2 \mu} e_{jk}^* g_{ijk}^{2D}(\vec{n}) \label{eq:dep2D}
\end{equation}
where $S$ is the area of the inclusion. The function $g_{ijk}^{2D}
(\vec{n})$ depends only on the direction $\vec{n}$ (see
Eq.~\ref{eq:dep2D_A2}). The explicit determination of the solution
depends on the expression of $e_{ij}^*$. In the present study, we
restrict to the case of strain tensors characterizing the plastic
rearrangement coaxial to the uniform imposed stress tensor
$\mathbf{\sigma^{\infty}}$, i.e.:
\begin{equation}
\mathbf{e}^* = \left(
\begin{array}{cc}
e^*_{xx} & 0\\
0 & e^*_{yy}
\end{array}
\right) \label{eq:estar2D}
\end{equation}
The geometry of the loading motivates this hypothesis. Furthermore,
Dasgupta \emph{et al.}~\cite{Dasgupta2013a,Dasgupta2013b} have shown
that this orientation minimizes the energy of interaction between the
external strain field and the inclusion at high enough strain. The
components of the strain tensor $\mathbf{\tilde{e}}$ induced by the
rearrangement in the surrounding matrix can then be computed
explicitly (see Appendix~\ref{sec:Ap_2D}).

In a biaxial experiment, the crucial quantity which characterizes the
shear and which governs the response of the material is the deviatoric
stress $\sigma_{xx} - \sigma_{yy}$. The significant component of the
stress redistribution to determine is thus the redistributed
deviatoric stress $\tsx - \tsy$, which will add to the imposed stress
field and will thus modify locally the total deviatoric stress. Here,
we obtain from \eqref{eq:dep2D} (see Appendix~\ref{sec:Ap_2D}):
\begin{equation}
\tsx - \tsy \propto \frac{S}{\pi r^2} \frac{\lambda + \mu}{\lambda
  + 2 \mu} f(\theta)
\label{eq:tsxtsy}
\end{equation}
with
\begin{equation}
  f(\theta) = - (\exs + \eys) \cos (2 \theta) - (\exs - \eys) \cos (4
  \theta)
\label{eq:f_2D}
\end{equation}
This function has a quadrupolar part which originates from the shear
part of the strain: $(\exs - \eys)$ and a bipolar part due to the
volumetric strain $(\exs + \eys)$. The function $f(\theta)$ is
independent of the elasticity coefficients because we are in two
dimensions. In three dimension the corresponding function depends on
the Poisson's ratio.

\begin{figure}[htbp]
\includegraphics[width=\columnwidth]{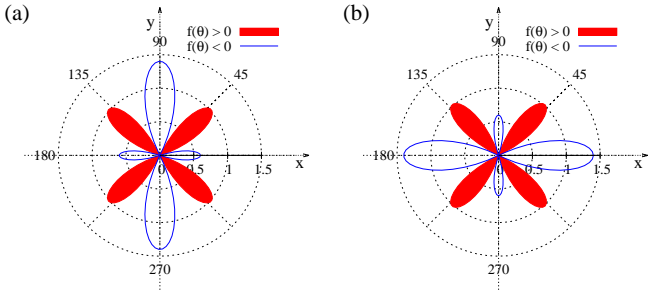}
\caption{Polar representation of the function $f(\theta)$
  (Eq.~\ref{eq:f_2D}). (a) Contractant event: $\exs = .3$ and
  $\eys=-.7$. (b) Dilatant event: $\exs = .7$ and
  $\eys=-.3$.}
\label{fig:polaire2D}
\end{figure}

Figure~\ref{fig:polaire2D}(a) (resp. (b)) shows polar representations
of the function $f(\theta)$ in a contracting (resp. dilating)
case. We observe a dominant quadrupolar response with an increase of
the deviatoric stress along directions close to $45^{\circ}$ from the
principal stresses (red curve of Fig.~\ref{fig:polaire2D}). The change
of volume induces a small modification of this inclination as well as
an asymmetry in the negative part of deviatoric stress redistribution
(blue curves of Fig.~\ref{fig:polaire2D}). This negative contribution
counteract the macroscopic loading $\sigma_{xx}^{\infty} -
\sigma_{yy}^{\infty}$.

\subsection{Generating numerically an Eshelby inclusion}
\label{sec:make-Eshelby}

A careful reading of Sec.~\ref{sec:principle}, especially of
Eq.~(\ref{eq:uGreen2D}), shows that the stress redistribution due to the
inclusion is obtained by integrating Green function for stress over
the boundary of the inclusion.  In the simulations, therefore, we will
define a small inclusion, and then modify the stress at the boundary.
But since the simulation is discrete, we must apply this stress in a
discrete way.

We specify a perturbing stress $\bsigma^*$, and not a perturbing
strain $\bm{e}^*$ because in Eshelby's calculation, it is the stress
that is propagated into the material [see Eq.~(\ref{eq:uGreen2D})].
The perturbing stress is obtained from the perturbing strain, using
the elastic coefficients.  In our case, the elastic coefficients are
not known, so that a more direct comparison can be obtained by
directly imposing a stress.

Let us suppose that we want to estimate the stress in some region
$\mathbb{V}$. We have \cite{Alexander98}
\begin{equation}
\bm{\sigma} = \frac{1}{S}
	\sum_{\alpha\in\mathbb{A}} \mathbf{F}_\alpha \otimes \mathbf{r}_\alpha,
\label{eq:stress-Love}
\end{equation}
where $S=|\mathbb{V}|$ is the area (in two-dimensions) of the
region where we calculate the stress.  The variable $\alpha$ labels
the contacts and $\mathbb{A}$ is the set of contacts with at least one
grain within $\mathbb{V}$.  (A grain is in $\mathbb{V}$ if its center
is in $\mathbb{V}$).  The force between the grains is
$\mathbf{F}_\alpha$, $\mathbf{r}_\alpha$ is a vector along which the
force is transmitted inside $\mathbb{V}$, and $\otimes$ indicates a
tensor (dyadic) product.

\begin{figure}
\includegraphics[width=0.8\columnwidth]{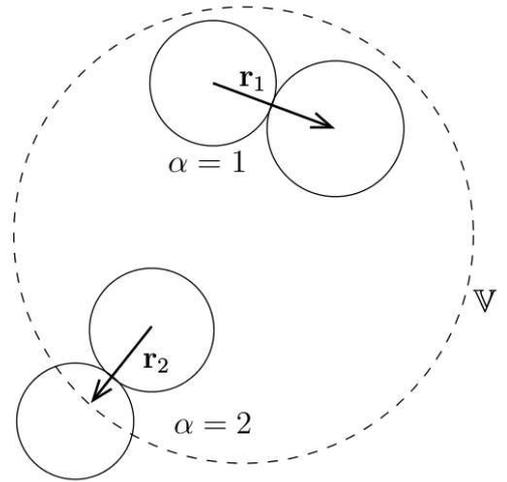}
\caption{Two contacts inside the region $\mathbb{V}$
where the stress is being calculated.
Upper contact ($\alpha=1$): both grains lie within $\mathbb{V}$,
$\mathbf{r}_1$ is the line of centers.
Lower contact ($\alpha=2$): only one grain lies within $\mathbb{V}$,
$\mathbf{r}_2$ extends from this grain, along the line of centers,
to the boundary of $\mathbb{V}$.}
\label{fig:contact-sketch}
\end{figure}

Some contacts in
$\mathbb{A}$ have both grains in $\mathbb{V}$ while others have only
one -- see Fig.~\ref{fig:contact-sketch}.  Let $\mathbb{B}$ be the set
of all contacts with two grains in $\mathbb{V}$ while $\mathbb{C} =
\mathbb{A} \backslash \mathbb{B}$ is the set of all contacts with one
grain in $\mathbb{V}$.  The vector $\mathbf{r}_\alpha$ depends on
whether the contact $\alpha$ is in $\mathbb{B}$ or $\mathbb{C}$.  If
$\alpha\in\mathbb{B}$ (both grains in $\mathbb{V}$), then
$\mathbf{r}_\alpha = \mathbf{x}_{j_\alpha} - \mathbf{x}_{i_\alpha}$,
where $i_\alpha, j_\alpha$ label the two particles participating in
contact $\alpha$.  On the other hand, if $\alpha\in\mathbb{C}$, then
$\mathbf{r}_\alpha = {\mathbf{x}^\circ_\alpha}-\mathbf{x}_{i_\alpha}$,
where $\mathbf{x}^\circ_\alpha$ is the point where the line of centers
crosses the boundary of $\mathbb{V}$.  We can therefore rewrite
Eq.~(\ref{eq:stress-Love}) as
\begin{equation}
\bm{\sigma} = \frac{1}{S}
	\sum_{\alpha\in\mathbb{B}} \mathbf{F}_\alpha \otimes
		(\mathbf{x}_{j_\alpha} - \mathbf{x}_{i_\alpha}) +
	\frac{1}{S}
	\sum_{\alpha\in\mathbb{C}} \mathbf{F}_\alpha \otimes
		(\mathbf{x}^\circ_\alpha-\mathbf{x}_{i_\alpha}).
\end{equation}
Instead of a sum over contacts, we can now write the stress as a sum
over grains.  We must define two new sets: let $\mathbb{G}$ be the set
of all grains whose centers are in $\mathbb{V}$, and let
$\mathbb{A}_i$ be the set of all contacts involving grain $i$.  Then we
have
\begin{equation}
\bm{\sigma} = \frac{1}{S}
	\sum_{i\in\mathbb{G}}
		\left[\sum_{\alpha\in\mathbb{A}_i} \chi_{i\alpha}
	\mathbf{F}_\alpha\right] \otimes \mathbf{x}_i
	+ \frac{1}{S}
		\sum_{\alpha\in\mathbb{C}} \mathbf{F}_\alpha \otimes \mathbf{x}^\circ_\alpha.
\label{eq:Love-grains}
\end{equation}
Here, $\chi_{i\alpha}=\pm 1$, depending on whether we have
$i=i_\alpha$ or $i=j_\alpha$.

Now the sum in brackets in Eq.~(\ref{eq:Love-grains}) vanishes when
the grains are in equilibrium, for it is simply the sum of forces
exerted on grain $i$.  The first term of Eq.~(\ref{eq:Love-grains})
thus vanishes, leaving only those contacts that cross the boundary:
\begin{equation}
\bm{\sigma} = \frac{1}{S}
	\sum_{\alpha\in\mathbb{C}} \mathbf{F}_\alpha\otimes\mathbf{x}^\circ_\alpha.
\label{eq:Love-boundary}
\end{equation}
This equation tells us that the stress inside a region is determined
just by the contacts that span the boundary.

Up to now, we have been considering these equations from a diagnostic
point of view, that is, the contact forces and grain positions are
extracted from numerical data, and then the stress tensor is
calculated in view of passing over to a continuum description.  Now we
will take a different perspective: the stress tensor on the left hand
side is a perturbing stress that we want to apply inside $\mathbb{V}$,
while the forces on the right hand side are modifications of the
contact forces that we will introduce in order to apply that stress.
Accordingly, we now write
\begin{equation}
\bm{\sigma}^* = \frac{1}{S}
	\sum_{\alpha\in\mathbb{C}} \mathbf{F}^*_\alpha
		\otimes\mathbf{x}^\circ_\alpha.
\label{eq:Love-impose}
\end{equation}
where we have written a star next to the stress tensor [as in
  Eq.~(\ref{eq:uGreen2D})], and next to the forces, to indicate this
new perspective.

Our method for generating an active inclusion consists of first
choosing the stress $\bsigma^*$, and then choosing forces
$\mathbf{F}^*_\alpha$ so that Eq.~(\ref{eq:Love-impose}) holds.  The
forces can then be added directly to the simuation.

We emphasize that our method begins with $\bsigma^*$, that appears in
Eq.~(\ref{eq:uGreen2D}), at the beginning of Eshelby's calcution.  The
alternative of imposing displacements~\cite{Puosi2014,Priezjev2015}
corresponds to imposing $\tui$ inside the inclusion.  But Eshelby's
calculation does not begin with $\tui$, but rather with $\bsigma^*$
(or $\bm{e}^*$ if the elastic moduli are known).

Now let us return to the description of our method and
consider the special case where $\mathbb{V}$ is a circle of
radius $R$ ($S=\pi R^2$), as shown in Fig.~\ref{fig:contact-sketch}.  The points
$\mathbf{x}_\alpha^\circ$ are all located on the circle.
Placing the origin at the center of the circle, their
coordinates can be written in terms of the $R$ and the unit vector
$\mathbf{n}_\alpha=\begin{pmatrix}\cos\theta_\alpha
\\ \sin\theta_\alpha \end{pmatrix}$.  We have
\begin{equation}
\bm{\sigma}^* = \frac{1}{\pi R}
	\sum_{\alpha\in\mathbb{C}} \mathbf{F}^*_\alpha
		\otimes \mathbf{n}_\alpha
\label{eq:Love-circle}
\end{equation}
When we specify $\bm{\sigma}^*$, Eq.~(\ref{eq:Love-circle}) contains
four equations and $2|\mathbb{C}|$ unknowns because each
$\mathbf{F}^*_\alpha$ is unknown and contains two components.  The
volume $\mathbb{V}$ is chosen to approximate the circular inclusion of
the Eshelby calculation, and therefore must be larger than the grain
size, leading to $|\mathbb{C}|\gg 1$.  Thus the system
(\ref{eq:Love-circle}) is always under-determined.  To add extra
conditions, we minimize $\sum_{\alpha\in\mathbb{C}}
|\mathbf{F}^*_\alpha|^2$.  The meaning of this condition is that we
look for the microscopic force distribution obeying
Eq.~(\ref{eq:Love-circle}) which disturbs the medium as little as
possible.  We therefore use Lagrange multipliers to minimize
$\sum_{\alpha\in\mathbb{C}} |\mathbf{F}^*_\alpha|^2$ subject to the
condition in Eq.~(\ref{eq:Love-circle}). We thus minimize
\begin{equation}
\mathcal{L} = \sum_{\alpha\in\mathbb{C}} |\mathbf{F}^*_\alpha|^2
	- \bm{\lambda :}
\left[ \pi R \bm{\sigma}^* - \sum_{\alpha\in\mathbb{C}} \mathbf{F}^*_\alpha
	\otimes \mathbf{n}_\alpha
\right].
\label{eq:grosLagrangian}
\end{equation}
Here, $\bm{\lambda}$ is matrix of Lagrange multipliers,
and the two points signify a total contraction between two matrices
($\mathbf{A:B} = A_{ij}B_{ij}$).

Differentiating Eq.~(\ref{eq:grosLagrangian}) by the forces,
we get
\begin{equation}
\mathbf{F}^*_\alpha = \frac{1}{2} \bm{\lambda} \mathbf{n}_\alpha.
\label{eq:inclusiondF}
\end{equation}
The Lagrange multipliers can be found by putting
the forces in to the conditions Eq.~(\ref{eq:Love-impose}):
\begin{equation}
2\pi r \bm{\sigma}^* = \bm{\lambda}
	\begin{pmatrix} A & C \\ C & B \end{pmatrix}.
\end{equation}
where
$A = \sum_{\alpha\in\mathbb{C}} \cos^2\theta_\alpha$,
$B = \sum_{\alpha\in\mathbb{C}} \sin^2\theta_\alpha$,
$C = \sum_{\alpha\in\mathbb{C}} \cos\theta_\alpha \sin\theta_\alpha$.

Before passing on to the results,
we remark that a pair of grains can be considered as a
contact in the set $\mathbb{C}$
even if the two grains are not touching.
In that case the contact force vanishes,
but that contact can be used to apply the perturbing stress.

\section{Results of the numerical experiment}
\label{sec:results}
In this section we study the response of the granular material to a
provoked Eshelby inclusion.

\subsection{Generation of a response to an Eshelby inclusion }

We compute the response to an Eshelby inclusion in the following
way. We first perform a compression of the granular material as
described in~\ref{sec:setup} up to a given deformation. We then
identify a small circle region of radius $R=L_0/80$ (approximately 6
grain radii) at the center of the sample.  This small circle is the
region $\mathbb{V}$ discussed above. For a region two times larger,
the results presented below are unchanged. We also tested $R=L_0/160$,
and the main effect seems to be to rotate the far-field response by of
order $10^\circ$.  If $R$ is reduced another factor of two, there are
not enough contacts to apply the perturbuation.

We then perform two simulations.  In the first simulation, we simply
continue the biaxial test without adding the perturbing force.  The
second simulation has exactly the same duration as the first, except
that the perturbing forces are gradually ``turned on'' over a time
long enough to avoid the generation of shock waves.

Then the stress $\bsigma_i$ on each grain $i$ is calculated as a sum over
that grain's contacts.
Using the notation of the previous section,
\begin{equation}
\bsigma_i = \frac{1}{\pi r_i^2}\sum_{\alpha\in\mathbb{A}_i}
	\chi_{i\alpha} \left(\mathbf{F}_\alpha \otimes \mathbf{r}_\alpha\right)
	\frac{r_i}{|\mathbf{r}_\alpha|}.
\end{equation}
where $r_i$ is the radius of grain $i$.
This is an application of Eq.~(\ref{eq:stress-Love}) to a single grain.
The last factor arises because only part of the line joining the centers
is inside the grain.

Since we have two simulations,
each grain has two stress values.
Let $\bsigma_i^{(0)}$ be the stress on grain $i$ in the unperturbed simulation,
and $\bsigma_i^{(1)}$ be the stress in the perturbed simulation.
We then calculate
\begin{equation}
\tilde\bsigma_i = \bsigma_i^{(1)} - \bsigma_i^{(0)}.
\end{equation}
Using this method,
we calculate a perturbation stress for each grain
that we can then use to visualize the effect of the inclusion.

\subsection{Description of the stress response}

\begin{figure*}
\begin{tabular}{ccc}
beginning & maximum density & maximum stress\\
\includegraphics[width=0.63\columnwidth]{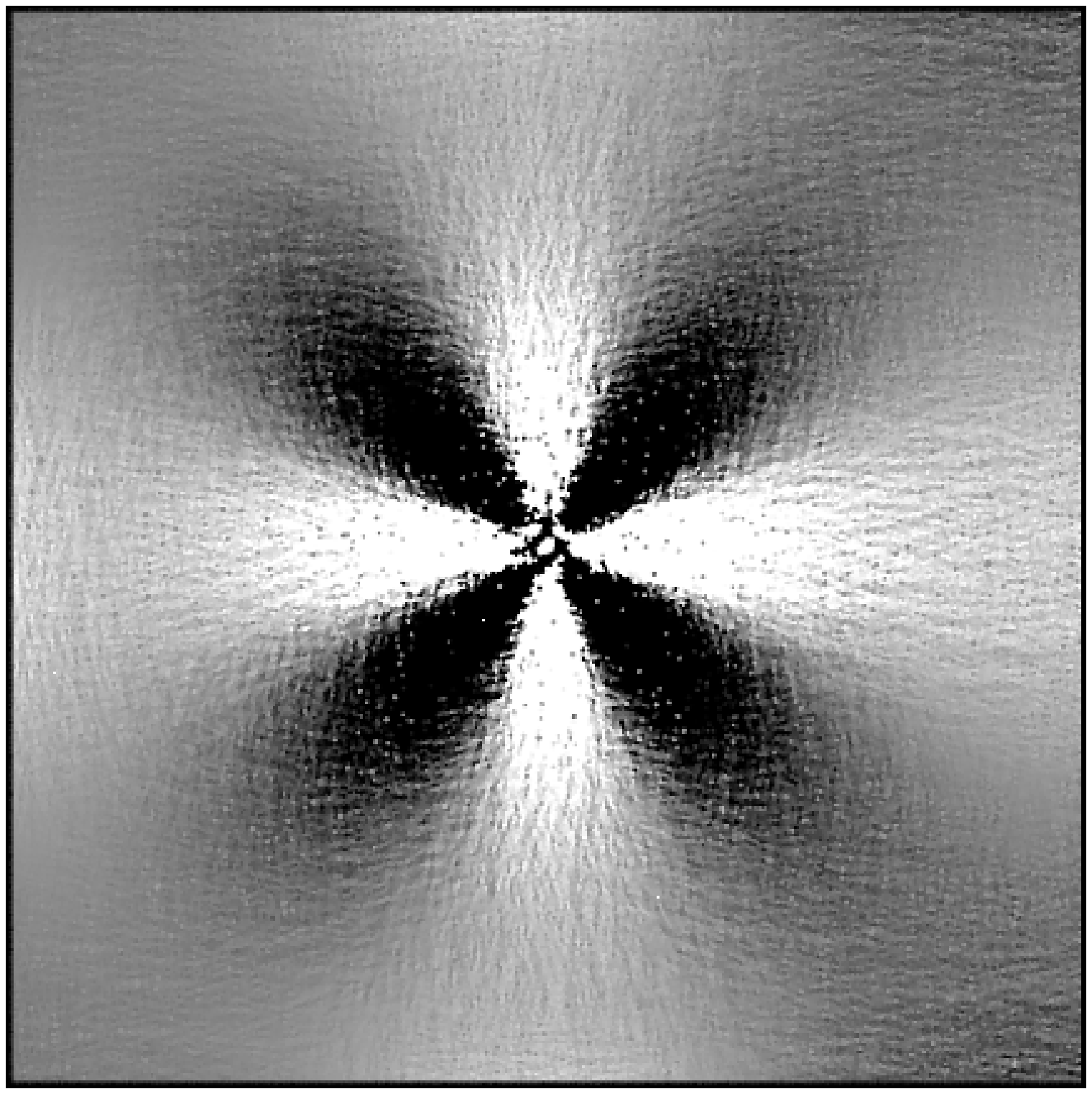}&
\includegraphics[width=0.63\columnwidth]{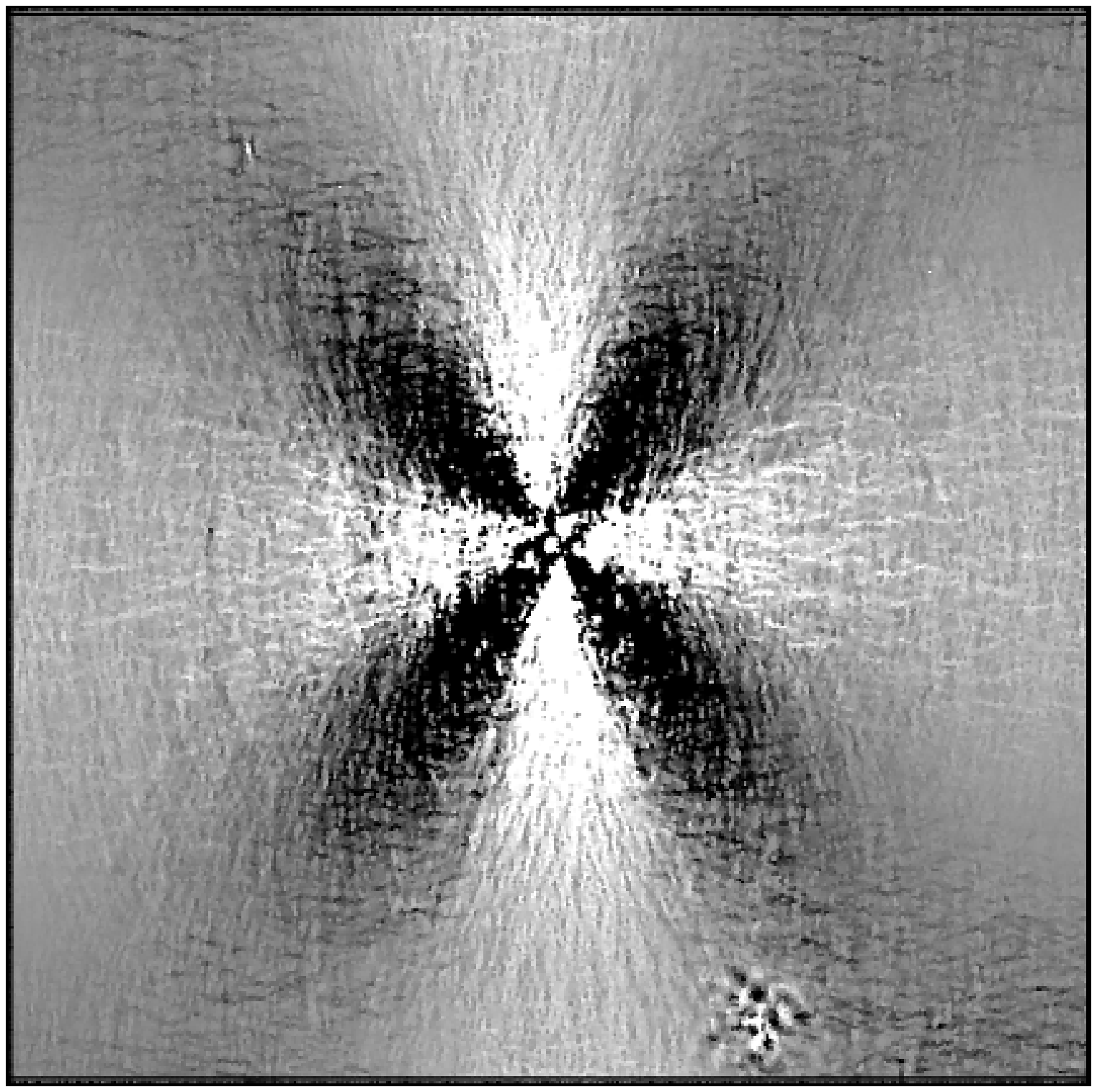}&
\includegraphics[width=0.63\columnwidth]{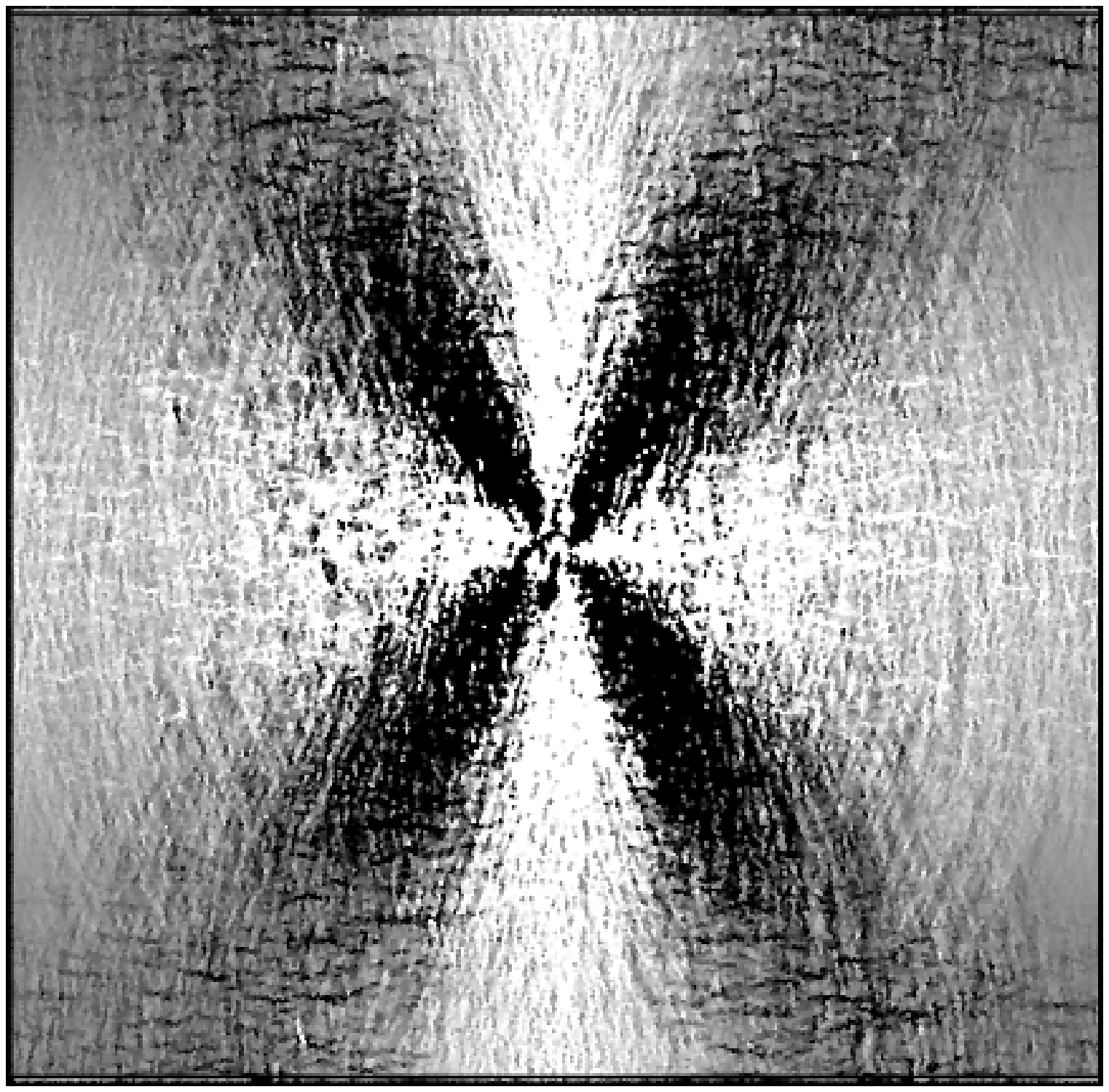}\\
$\varepsilon_{yy} = 0.001\%$ & $\varepsilon_{yy} = 0.505\%$ & $\varepsilon_{yy} = 0.900\%$\\
\includegraphics[width=0.63\columnwidth]{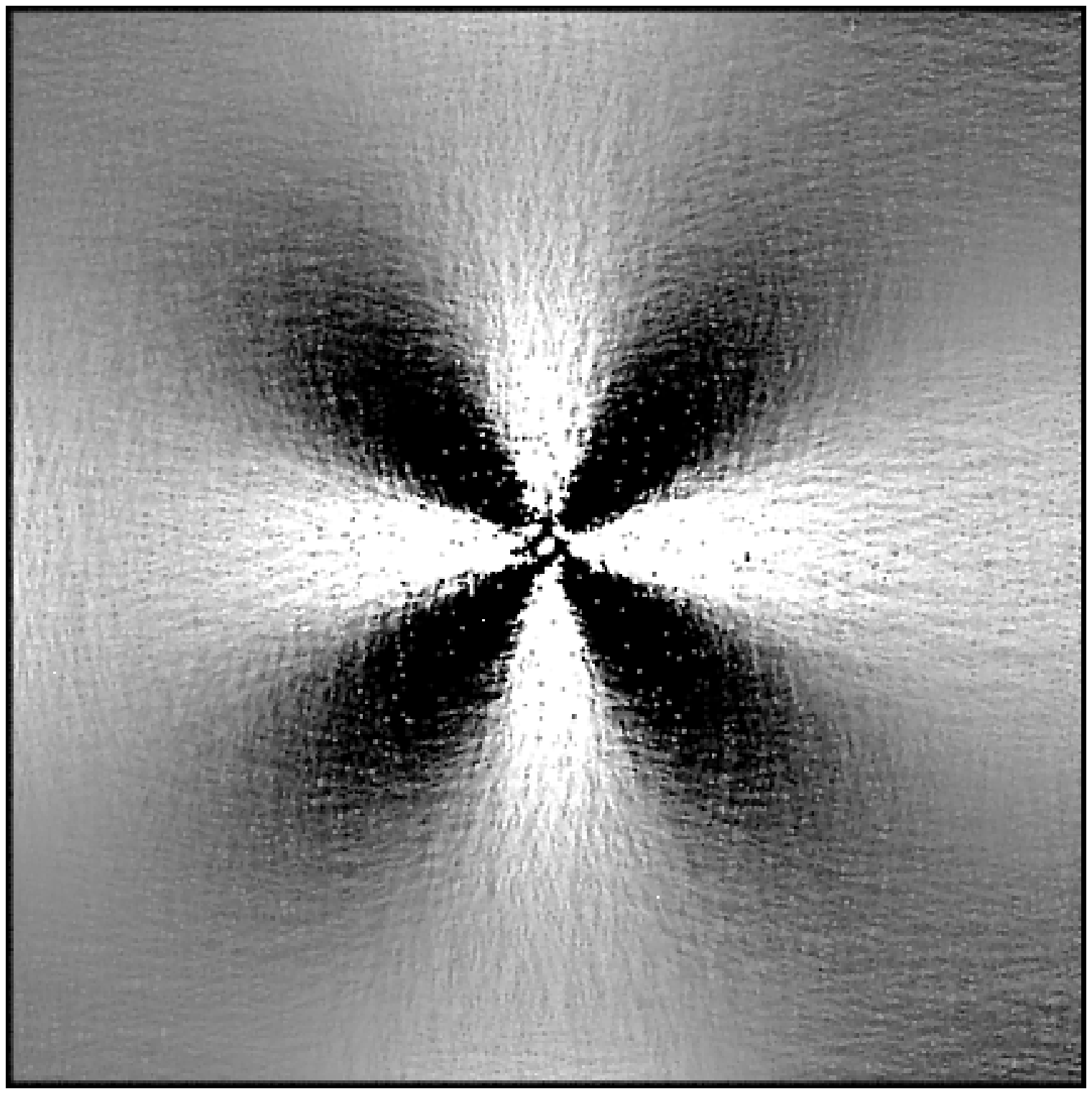}&
\includegraphics[width=0.63\columnwidth]{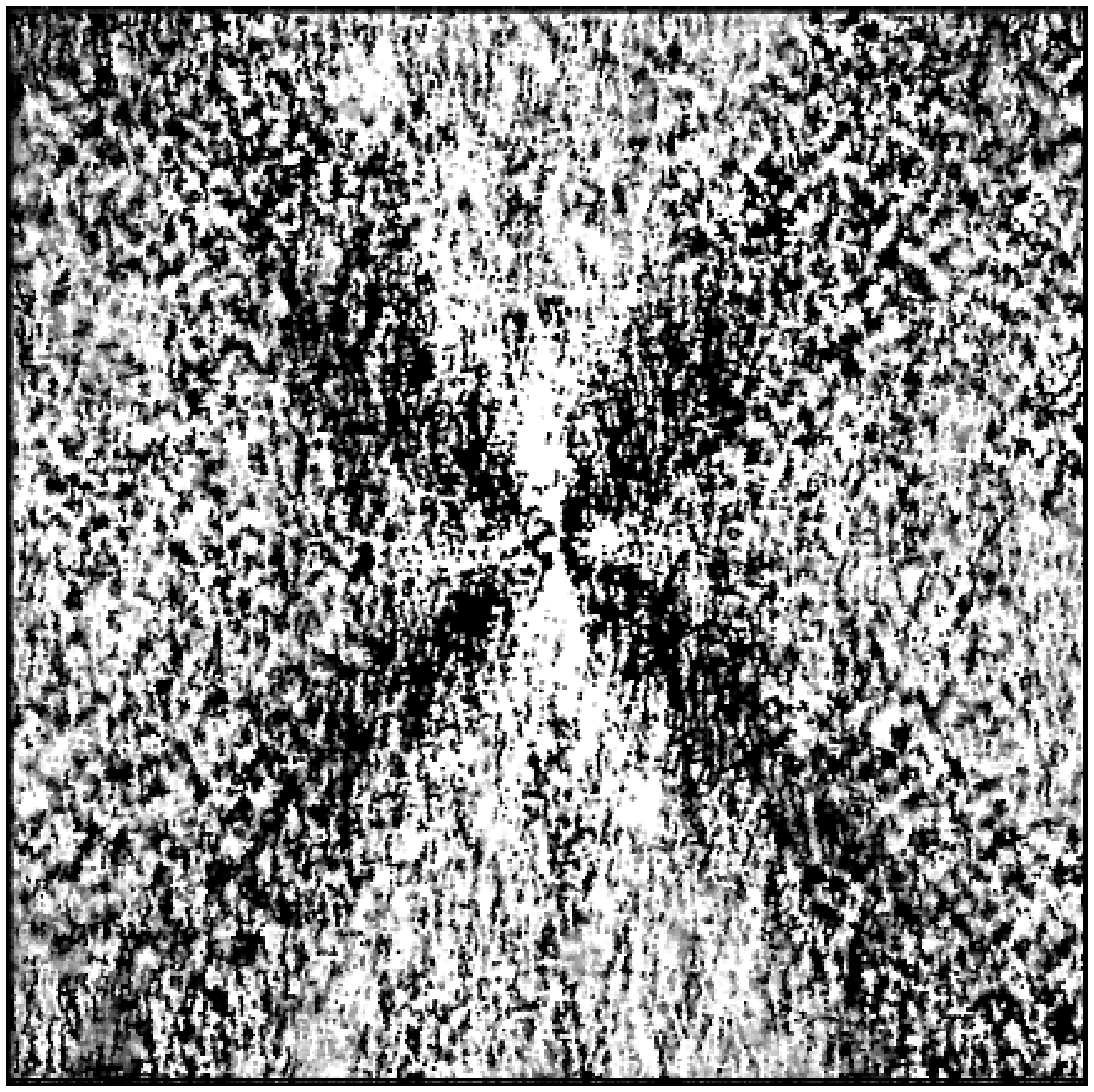}&
\includegraphics[width=0.63\columnwidth]{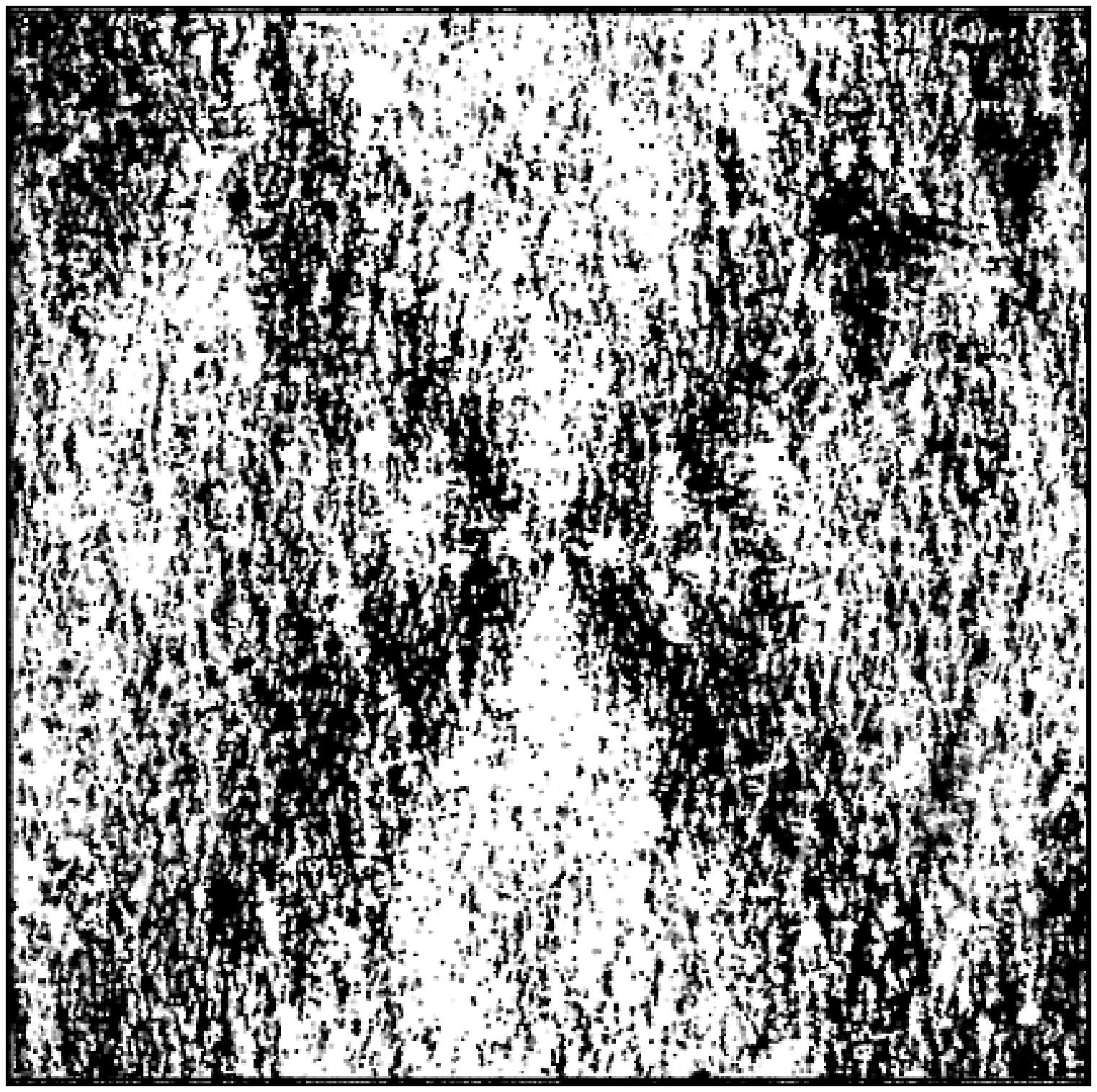}\\
$\varepsilon_{yy} = 0.001\%$ & $\varepsilon_{yy} = 0.239\%$ & $\varepsilon_{yy} = 0.405\%$\\
\includegraphics[width=0.63\columnwidth]{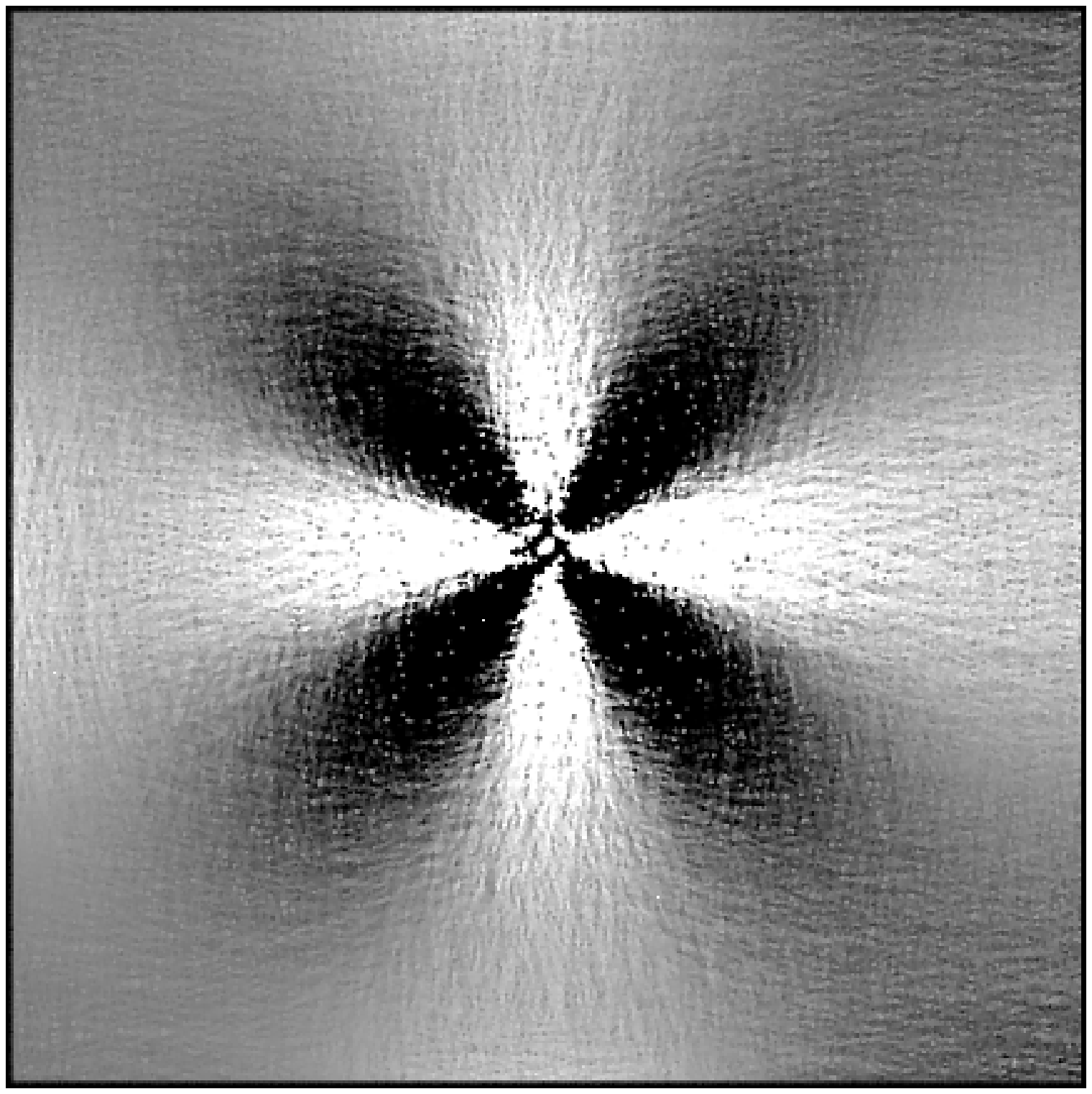}&
\includegraphics[width=0.63\columnwidth]{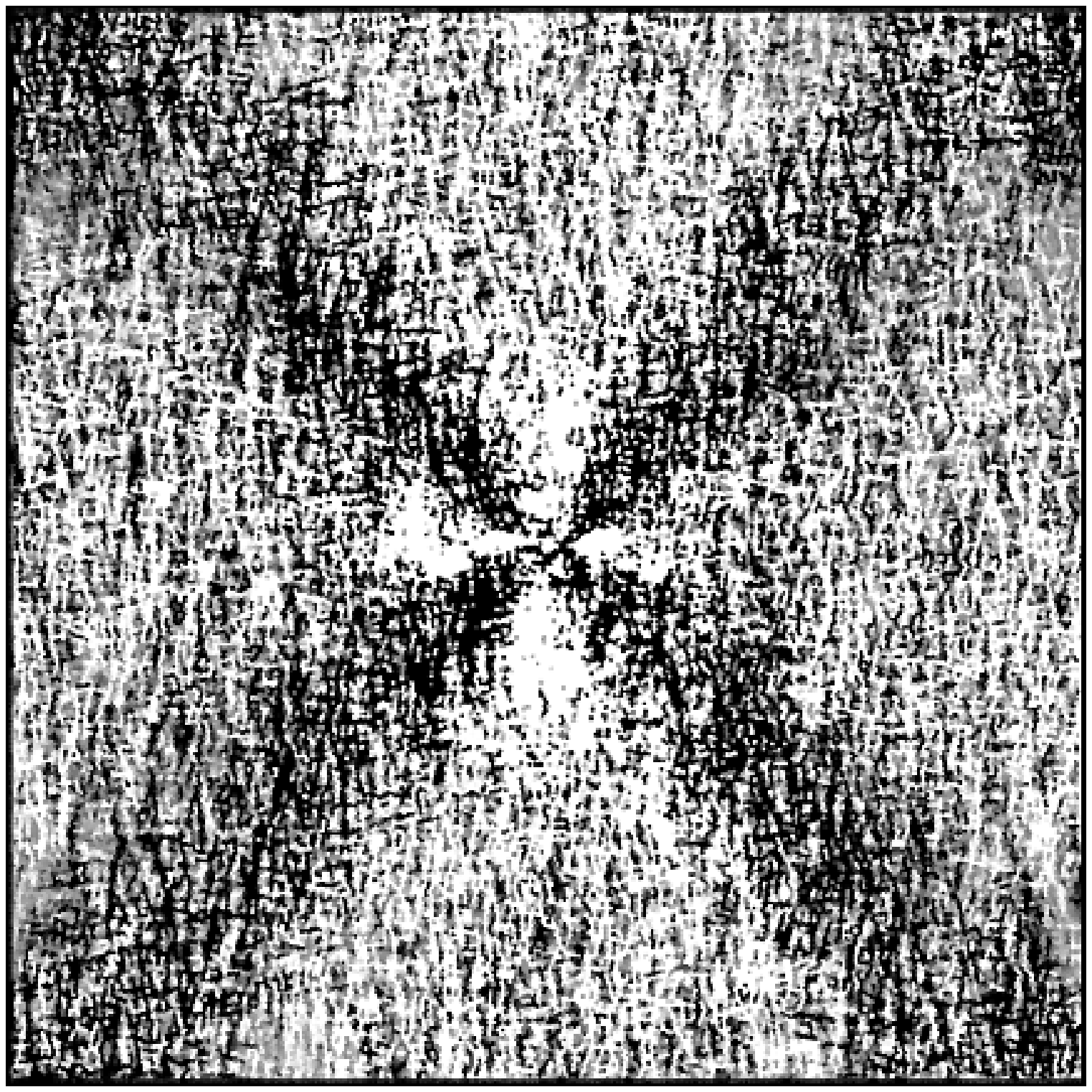}&
\includegraphics[width=0.63\columnwidth]{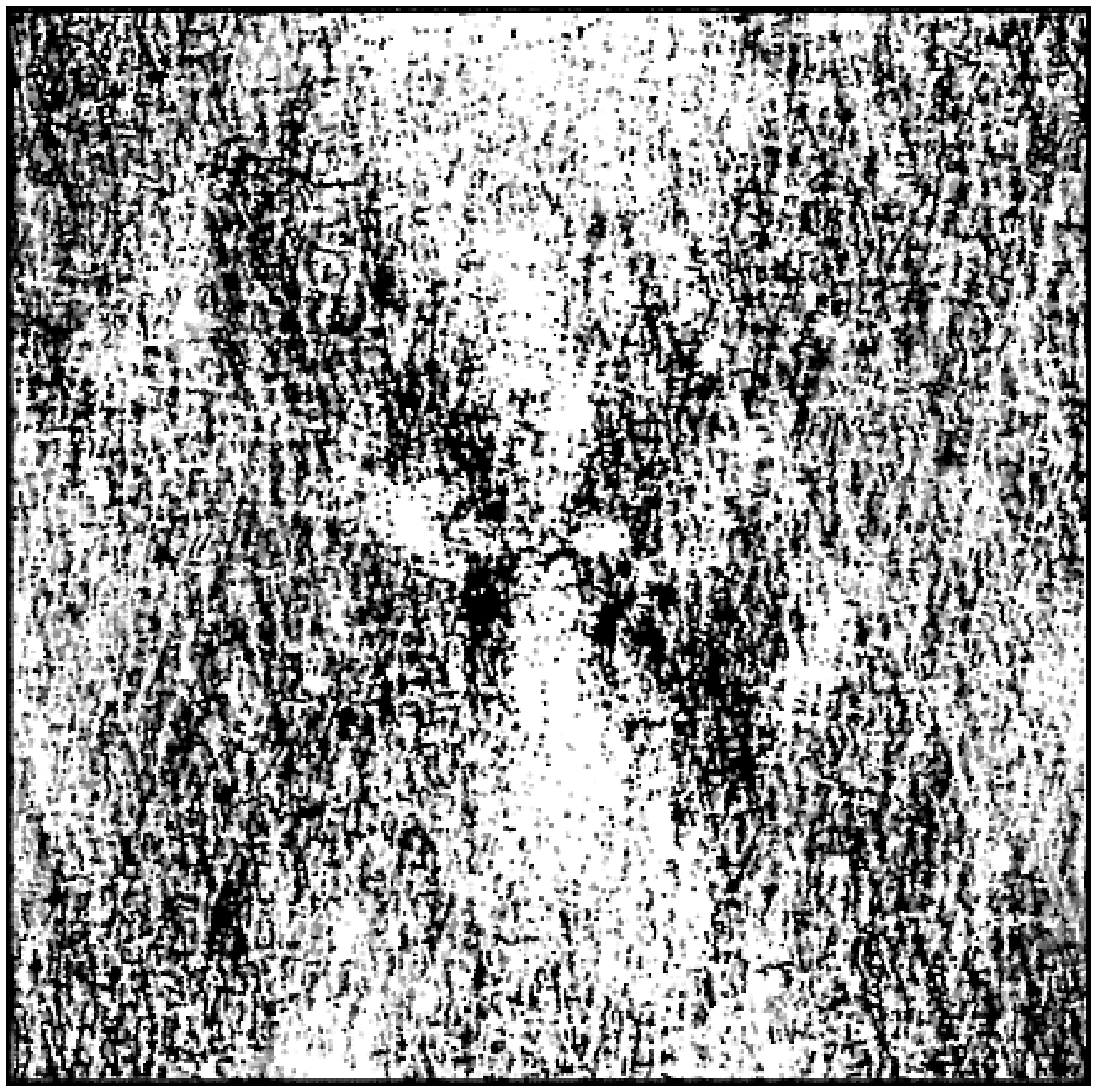}\\
$\varepsilon_{yy} = 0.001\%$ & $\varepsilon_{yy} = 0.142\%$ & $\varepsilon_{yy} = 0.404\%$
\end{tabular}
\caption{Examples of generated Eshelbies, $N=256\times 256$ grains.
  Top row: $\mu_m=2$, middle row: $\mu_m=0.6$, bottom row $\mu_m=0.2$.
  $\bsigma^*$ is coaxial with the load; white grains correspond to a
  reduction in deviatoric stress.  The value of the global strain is
  given under each panel.  }
\label{fig:Eshelby-example}
\end{figure*}

We now focus on the spatial distribution of {$\tilde\sigma_{yy,i} -
  \tilde\sigma_{xx,i}$} inside the granular sample. In
Fig.~\ref{fig:Eshelby-example}, we show the change in deviatoric
stress {$\tilde\sigma_{yy,i} - \tilde\sigma_{xx,i}$}. The map is
obtained by shading each grain accordingly to the value of the
deviatoric stress. The perturbing stress applied in
Fig.~\ref{fig:Eshelby-example} is $\bsigma^* = \alpha^*
p_0 \begin{pmatrix} 1 & 0 \\ 0 & -1 \end{pmatrix}$.  Here, $\alpha^*$
is a dimensionless number that measures the strength of the
perturbation with respect to the confining pressure $p_0$.  Throughout
this paper, $\alpha^*=10^{-2}$.  Some tests (not shown) were done with
$\alpha^*=10^{-3}$, and the resulting response is simply divided by
10, indicating that we obtain a linear response.

We considered the biaxial tests of Figs.~\ref{fig:stress-strain} et~\ref{fig:volume-strain}. For each test, the stress response at three instants are shown:
one at the beginning of the test,
one near the maximum density (minimum of the curves shown in Fig.~\ref{fig:volume-strain}),
and one near the maximum load.

At the beginning of the test, all three tests show a clear, four-lobed
pattern characteristic of the Eshelby stress tensor. The dark, maximum
stress lobes are oriented at $45^\circ$ from the horizontal. After
this initial loading, in the case of $\mu_m=2$ simulation, the Eshelby
pattern remains when the axial strain increases, but the angle of the
bands increases. Meanwhile, the other two tests ($\mu_m=0.2$ and $0.6$)
show a quite different behavior: the band angle remains near
$45^\circ$, but the texture of the image becomes much more grainy. We
will discuss and interpret those different behaviors in
section~\ref{sec:discussion}.

\subsection{Characterization of the angular response} \label{subsec:angles}
\begin{figure}
\includegraphics[width=\columnwidth]{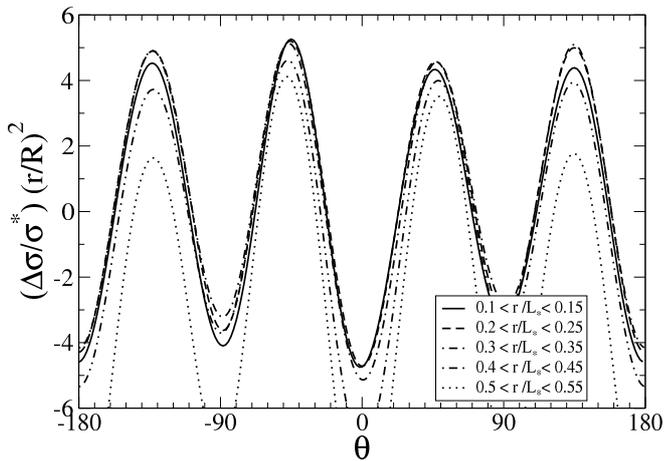}
\caption{An estimate of the angular dependence of the perturbation stress
$f(\theta)$. The stress on grains is averaged over radial bins and
then rescaled to esimate $f(\theta)$.
This example is for $\mu_m=0.6$, $\varepsilon_{yy}=10^{-5}$.}
\label{fig:Eshelby-r}
\end{figure}

We obtain quantitative information from Fig.~\ref{fig:Eshelby-example}
about the function $f(\theta)$. We divide the grains into classes
according to their distance $r$ from the center of $\mathbb{V}$, and
then into subclasses according to their angle $\theta$. An average
value of $\Delta\sigma_i$ is calculated for each subclass. Then this
average is multiplied by the $r^2$. This allow us to test the $1/r^2$
scaling of the redistributed stress.  The resulting function for
different classes of $r$ is shown in Fig.~\ref{fig:Eshelby-r}.  We
observe a good collapse close to the perturbation ($r\le 0.35L_0$)
showing that the far field approximation (leading to keep only the
order $1/r^2$ for the stress field) is sufficient to capture the
radial response.  Only the curves corresponding to the largest $r$
differs significantly from the others, certainly because sampled
region approaches the boundary of the simulation. (The simulation is
approximately square with side length a bit larger than $1.1L_0$.)

The scaling in Fig.~\ref{fig:Eshelby-r} is based on Eq.~(\ref{eq:tsxtsy}).
This equation shows that the stress response should be proportional
to $S/r^2$, where $S$ is the area of the inclusion ($S=\pi R^2$ in our case).
Dividing by $R^2/r^2$ removes this dependence.
Futhermore, the response is proportional to the perturbing stress,
so we divide by $\sigma^*$ yielding a function $f(\theta)$ of order unity.

We want to identify the position of the maxima of $f(\theta)$.  To
this end, we consider the grains with $0.1 < r/L_0 < 0.3$, and sort
them into classes according to their angle $\theta$, and calculate the
average change in stress in each class to obtain an estimate of
$f(\theta)$.  This function is then approximated by a Fourier series
\begin{equation}
f(\theta) \approx a_0 + \sum_{k=1}^n a_k \sin k\theta +
	\sum_{k=1}^n b_k \cos k\theta,
\label{eq:fit}
\end{equation}
with $n=8$.
The maxima of this Fourier series are identified as the angle of the
diagonal bands of the Eshelby.

We study the band angle as a function of loading. As the loading
increases, a new problem appears: the packing becomes fragile, and the
inclusion can trigger other events throughout the packing that swamp
the change in stress caused directly by the inclusion.  To get around
this problem, we search for times where the packing is relatively
stable (absence of kinetic energy peaks), and furthermore, we reduce
the strain rate by a factor of 100, so that we will have time to
``turn on'' the forces in the inclusion.
\begin{figure}
\includegraphics[width=\columnwidth]{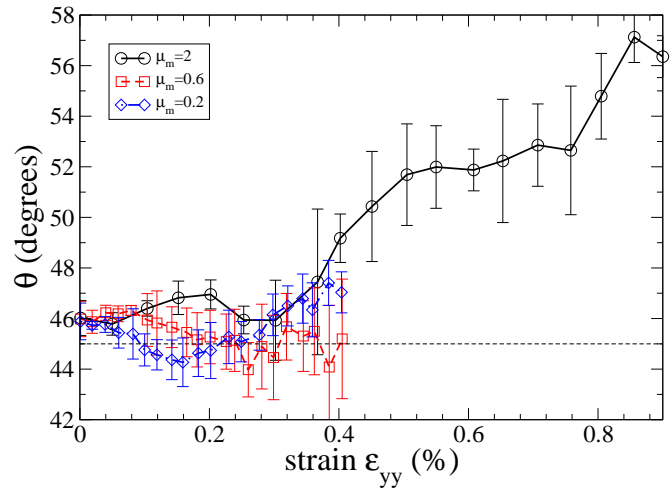}
\caption{Angle of diagonal bands for simulations with different
friction coefficients.
For each value of the strain,
four angles are averaged together to obtain a mean.
The error bars give an idea of the scatter of the different angles measured.
The horizontal line indicates an angle of $45^\circ$.
}
\label{fig:Eshelby-angles}
\end{figure}

In Fig.~\ref{fig:Eshelby-angles}, we show the angle found in this way
for the three biaxial tests of Figs.~\ref{fig:stress-strain}
and~\ref{fig:volume-strain}.  For each strain value, there are four
angles, one for each of the diagonal bands giving its angle relative
to the horizontal.  These angles are averaged together to obtain a
point on the graph.  The error bars estimate the scatter of the
points; they show the uncertainty of the mean. We observe that at the
beginning of the loading all the angles are close to $45^\circ$ but
with some fluctuations around this value. When the loading increase,
the departure from $45^\circ$ becomes important. We will further
discuss this figure in the next section.

To conclude, the mental construction of Eshelby underlying his
computation can be implemented in a numerical simulation of dry
frictional disks. In spite of the discrete nature of the material, the
continuum mechanics solutions are clearly relevant as the observed
response is very similar to the one predicted analytically. This
picture is modified when the granular material if further loaded and
the friction coefficient decreased.

\section{Discussion}
\label{sec:discussion}
In this part we discuss the departure from the quadrupolar response
observed in Figure~\ref{fig:Eshelby-example} either due to the value
of the friction coefficient or to the loading of the material. As
shown in the previous part, for an isotropically slightly
pre-compressed granular material (initial preparation), the response
of the material to a localized shear event is compatible with an
elastic response (first column of
Fig.~\ref{fig:Eshelby-example}). This picture is modified in two
different ways when the material is submitted to an increasing
deviatoric stress depending on the value of the friction
coefficient. We discuss the different kind of departures from this
ideal response in the following.

\subsection{Beginning of the loading}
As can be seen on the first column of
figure~\ref{fig:Eshelby-example}, at the very beginning of the loading
the response is very close to the theoretical prediction recalled in
section~\ref{subsec:2D} and in
Figure~\ref{fig:polaire2D}. Quantitatively, as discussed in the
preceding section, the radial response is in $1/r^2$ and the angular
response is dominated by a quadrupolar dependence
(Figure~\ref{fig:Eshelby-r}),
with a maximum stress redistribution close to $45^\circ$.
The direct observation of the stress
fields shown in figure~\ref{fig:Eshelby-example} shows nevertheless
some differences with the response of a continuous material: we can
observe that a filamentary pattern underly the smooth
response. Consequently, the observed stress redistribution is also
compatible with a description of the granular material in term of
force chains. The observed striations could indeed indicate that
stress is concentrated on grains in force chains. Nevertheless, on
a large scale, a smooth quadrupolar response is observed.

As the loading increases, two different behaviors are observed.
For $\mu_m=2$ (top row), the texture of the patterns remains
the same, but the angles depart significantly from $45^\circ$.
For $\mu_m=0.6$ and $\mu_m=0.2$,
the angles remain close to $45^\circ$,
but the texture changes significantly.
We now discuss these changes.

\subsection{Further loading: High value of $\mu_m$}
When the friction coefficient is artificially high to prevent sliding
contacts (upper line of Figure~\ref{fig:Eshelby-example}), we observe
a clear modification of the angles characterizing the stress
resdistribution. Meanwhile, the increase in the fluctuations is much
smaller than for the smaller friction coefficients. A change of
inclination can still be interpreted in an Eshelby picture and we will
discuss two possible origins in the following: the effect of the
volumetric strain and the effect of anisotropical elasticity.

\subsubsection{Polar response and volumetric strain}
As discussed in section~\ref{subsec:2D}, the quadrupolar response due
to the shear part of the rearrangement is modified by the volumetric
contribution. To make explicit this dependence,
we study the polar function $f(\theta)$ of
Eq.~\ref{eq:f_2D}. The minima of the function are obtained for $\theta
= 0^{\circ} [90^{\circ}]$ and the maxima for
\begin{equation}
\cos 2 \theta_E = \frac{1}{4} \frac{e_{xx}^* + e_{yy}^*}{e_{yy}^*
  - e_{xx}^*} \label{eq:theta_E}
\end{equation} For the isovolumic case ($e_{xx}^* + e_{yy}^*=0$), we obtain $\theta_E
= 45^{\circ} [90^{\circ}]$.

Our active inclusions correspond to an imposed traceless stress tensor
$\text{Trace} (\bm{\sigma}^*) = 0$.
If the inclusion is made of an isotropic elastic material,
the associated strain tensor will also be traceless.
But the loading process introduces anisotropy into the contact
network and thus into the elastic constants,
as will be discussed in the next section.
In such cases, $\text{Trace} (\bm{e}^*)$ may be non-zero.

If we suppose that the inclusion changes its volume, a reasonable hypothesis is
that $\exs > 0$ and $\eys < 0$, i.e. that the local deformation follows
the overall macroscopic deformation. We then obtain for the range of
possible values $\theta_E \in [\frac{1}{2} \text{acos} \left(
  \frac{1}{4} \right) ; \frac{1}{2} \text{acos} \left( - \frac{1}{4}
  \right)]$, i.e. $\theta_E \in [37.8^{\circ};52.2^{\circ}]$. The
values greater than $45^{\circ}$ correspond to $\exs + \eys > 0$,
i.e. to dilating plastic events, while the values lesser than
$45^{\circ}$ correspond to contracting plastic events. Those two
cases are shown on Fig.~\ref{fig:polaire2D}: the angle
of the positive part of the deviatoric stress redistribution is indeed
smaller than $45^{\circ}$ in the case of a contracting event
(Fig.~\ref{fig:polaire2D}(a)) and larger in the case of a dilating one
(Fig.~\ref{fig:polaire2D}(b)).
The range of possible angles found here
correspond to the direction of the maximum of the redistributed
deviatoric stress from a single Eshelby inclusion. It differs from the
range obtained in~\cite{Ashwin2013}, where the range of directions
of coupled Eshelby inclusion is determined by minimizing the elastic energy
of interaction between the inclusions.

The result of this calculation is that only small departure from
$45^{\circ}$ can be explained by non-isovolumic transformation. It
could for example explain the fluctuations measured in
Figure~\ref{fig:Eshelby-angles} for $\mu_m = 0.2$ and $\mu_m = 0.6$ but
not the large increase observed in the $\mu_m = 2$ case.

\subsubsection{Anisotropy of the elastic matrix}
\begin{figure}
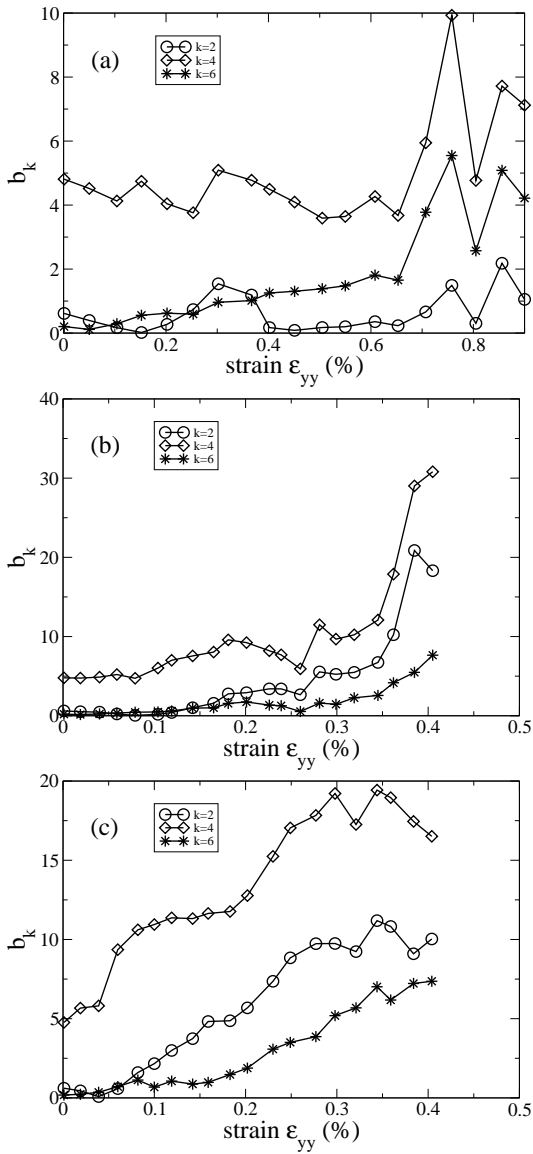

\centering
\includegraphics[width=0.8\columnwidth]{fig9a.eps}
\includegraphics[width=0.8\columnwidth]{fig9b.eps}
\includegraphics[width=0.8\columnwidth]{fig9c.eps}
\caption{Harmonic components obtained from the fits obtained using
  equation~\ref{eq:fit}. (a) $\mu_m = 2$, (b) $\mu_m = 0.6$ and (c) $\mu_m =
  0.2$.}
\label{fig:harmonic}
\end{figure}

In this part, we will neglect possible change of volume of the
inclusion ($\text{Trace} (\bm{e}^*) = 0$) but we will consider the
Eshelby problem in the case of anisotropic elasticity. Indeed,
  the loading process may induce some anisotropy of the elasticity
  of the material~\cite{Acoustics,LudingMM}.
The computation of the
Green's function in an anisotropic material is done in
Appendix~\ref{sec:Ap_aniso}. Equation~\ref{eq:aniso} shows that in the
case of an orthorhombic bidimensional material with a small
anisotropy, terms in $2\theta$ and $6\theta$ appear. On the
contrary, change of volume of the inclusion do not generate $6\theta$
terms. So the presence of $6\theta$ terms in $f(\theta)$ may be viewed as
an indication of anisotropic elasticity. This can be checked by the
Fourier analysis done in section~\ref{subsec:angles} (fit of the
measured response by equation~\ref{eq:fit}). The different components
of the fit are shown in Figure~\ref{fig:harmonic}. We observe that
indeed the $6\theta$ component is the second most important harmonic
in the case of $\mu_m=2$. This supports the hypothesis
that the observed modification of the angle is due to the effect of
material anisotropy, and not to volume change of the inclusion.

This effect is less clear for $\mu_m = 0.6$ and $\mu_m = 0.2$ where the
second dominant harmonic is $2\theta$ which can originate both
from volumetric effects and anisotropical effects. In fact, for those values of $\mu_m$, the principal effect
of the loading is not modification of the angles (that fluctuate close to
$45^{\circ}$  - see Figure~\ref{fig:Eshelby-angles}) but rather the
increase of fluctuations (see Figure~\ref{fig:Eshelby-example}) as
will be discussed in the next part.

\subsection{Further loading: Low value of $\mu_m$}
For $\mu_m = 0.6$ and $\mu_m = 0.2$ at intermediate and large values of the strain
(lower right four panels in Fig.~\ref{fig:Eshelby-example})
the filamentary striations visible in the other panels of Fig.~\ref{fig:Eshelby-example})
are replaced by a grainy strongly fluctuating texture.
In spite of this, the response at large scales remains quadrupolar, with angles close to $45^\circ$.

This change in texture cannot be
described in the Eshelby formalism, which supposes that the material
is continuous. We do not think that this difference of texture is due
to the disorder of the contacts between grains, but likely due to
sliding contacts. Indeed, the $\mu_m=0.2$ and $\mu_m=0.6$, there are
many sliding contacts (up to $35\%$ and $15\%$
of all contacts, respectively) but at $\mu_m=2$, sliding contacts remain relatively
rare (less than $2.5\%$ of all contacts).

The permanence of the quadrupolar structure at large scale shows than
an elastic description of a granular material can co-exist with
different mesoscopic behaviors: either force chains
($\varepsilon_{yy}=10^{-5}$ or $\mu_m=2$) or sliding contacts (large
$\varepsilon_{yy}$ and $\mu_m=0.2,0.6$).  When many contacts are
sliding, the granular matrix does not behave as an elastic
material. We then expect a stress redistribution which differs from
the elastic Eshelby response. However, at large scales, we still see
in Fig.~\ref{fig:Eshelby-example} the Eshelby structure which is a
characteristic of the material elasticity. It is unclear if it is
possible to quantify objectively such a separation of scales.

\section{Conclusion}
In this paper, we considered the effect of a local transformation on
the mechanical stress distribution. For this, we applied the
theoretical construction proposed by Eshelby to numerical
simulations where extra forces are applied on some grains inside a
granular material. In response to this local applied stress, the
material relaxes stress at large distance. This stress relaxation may
be, at least in some cases, successfully described by a theory where
the granular material is treated as a continuous and elastic
material. The presence of disorder of contact forces (such as the so
called force-chains) does not disrupt this elastic response, at least
at the macroscopic scale. If sliding contacts are supressed,
we show that the material remains elastic at the macroscopic scale,
even close to failure, but the elastic coefficients become anisotropic.

Our numerical simulations are performed with beads of finite
stiffness. The ratio $p_0/k_N=5.10^{-4}$ between the confining
pressure and the contact stiffness, is the typical value of the
relative deformations of the beads. It is interesting to compare it to
a real 3D experiment. If we consider glass beads following Hertz law
which are confined under a pressure $P = 100~kPa$, we may estimate a
relative deformation of the beads of $3.10^{-4}$. So the value of the
relative deformation is of the same order of magnitude as many biaxial
compression experiments, such as experiments where stress relaxations
along preferential directions are
observed~\cite{LeBouil2014,LeBouil2014b}.

Some departures from the elastic behavior are clearly visible in our
simulations. They may be linked to the occurrence of sliding
contacts. The departure from an elastic response of the matrix is an
interesting, but very challenging, problem. We may expect that the
granular skeleton behaves elastically for deformations of the
inclusion which are smaller than some plasticity threshold
$\varepsilon_{th}$. The variation of $\varepsilon_{th}$ with the
microscopic friction coefficient $\mu_m$, with the reduced stiffness
of the beads $p_0/k_N$, and with the proximity of the material failure are
still unknown. This will be the subject of a future work.

\appendix
\begin{widetext}
\section{Computation of the released stress} \label{sec:Ap_2D}

The bidimensional Green function is:
\begin{eqnarray}
G_{ij}^{(2D)}(\vec{r} - \vec{r'}) = \frac{1}{4 \pi \mu} \frac{\lambda
  + \mu}{\lambda + 2 \mu} \left[ - \frac{\lambda + 3 \mu}{\lambda +
    \mu} \delta_{ij} \ln |\vec{r}-\vec{r'}| + \frac{(x_i - x_i')(x_j -
    x_j')}{|\vec{r}-\vec{r'}|^2}\right]
\end{eqnarray}
where $\vec{r} = \left(
\begin{array}{c}
x\\
y
\end{array}
\right) = \left(
\begin{array}{c}
r \cos \theta\\
r \sin \theta
\end{array}
\right)$ are the coordinates in the bidimensional plane.

The displacement field in the matrix can be obtained by integrating
the Green function along the contour $C$ of the inclusion:
\begin{eqnarray*}
  \tui (\vec{r}) &=& \frac{1}{4 \pi \mu} \frac{\lambda + \mu}{\lambda
    + 2 \mu} \oint_C \left[ - \frac{\lambda + 3 \mu}{\lambda + \mu}
    \delta_{ij} \ln|\vec{r} - \vec{r'}| + \frac{(x_i - x_i')(x_j -
      x_j')}{|\vec{r} - \vec{r'}|^2}\right] \sigma_{jk}^* dc_k'\\ &=&
  \frac{\sigma_{jk}^*}{4 \pi \mu} \frac{\lambda + \mu}{\lambda + 2
    \mu} \iint_S \left[ \frac{\lambda + 3 \mu}{\lambda + \mu} n_k
    \delta_{ij} +\left( - \delta_{ik} n_j - \delta_{jk} n_i + 2 n_i
    n_j n_k \right)\right] \frac{dS'}{|\vec{r} - \vec{r'}|}
\end{eqnarray*}
Where $\vec{n} = \frac{\vec{r} - \vec{r'}}{|\vec{r} - \vec{r'}|}$.
Using $\sis = \lambda e_{ll}^* \delta_{ij} + 2\mu \eis$ one obtains:
\begin{equation}
  \tui (\vec{r}) = \frac{e_{jk}^*}{2\pi} \frac{\lambda + \mu}{\lambda
    + 2 \mu} \iint_S \left[ \frac{\mu}{\lambda + \mu} (\delta_{ij}
    n_k + \delta_{ik} n_j - \delta_{jk} n_i) + 2 n_i n_j n_k \right]
  \frac{dS'}{|\vec{r} - \vec{r'}|} \label{eq:dep2D_A1}
\end{equation}

Consequently, far from the inclusion, we can write:
\begin{equation}
\tilde{u_i}(\vec{r}) = \frac{S}{2 \pi r} \frac{\lambda +
  \mu}{\lambda + 2 \mu} e_{jk}^* g_{ijk}^{2D}(\vec{n}) \label{eq:dep2D_A2}
\end{equation}
with $g_{ijk}^{2D}(\vec{n}) = \frac{\mu}{\lambda + \mu} (\delta_{ij}
n_k + \delta_{ik} n_j - \delta_{jk} n_i) + 2 n_i n_j n_k$.

For the particular case when $\mathbf{e^*} = \left(
\begin{array}{cc}
e^*_{xx} & 0\\
0 & e^*_{yy}\\
\end{array}
\right)$ we have:
\begin{eqnarray*}
\ad{u}_{r} (\vec{r}) &=& \frac{S}{2\pi r} \frac{\lambda +
  \mu}{\lambda + 2 \mu} \left[
  \frac{\lambda + 2\mu}{\lambda + \mu} (e_{xx}^* - e_{yy}^*) \cos
  2\theta + (e_{xx}^* + e_{yy}^*)
  \right]\\
\ad{u}_{\theta} (\vec{r}) &=& \frac{S}{2\pi r} \frac{\lambda +
  \mu}{\lambda + 2 \mu} \left[
  -\frac{\mu}{\lambda + \mu} (e_{xx}^* - e_{yy}^*) \sin 2\theta \right]
\end{eqnarray*}

The components of the strain tensor in polar coordinates are then:
\begin{eqnarray*}
\ad{e}_{r r} &=& \frac{S}{2\pi r^2} \frac{\lambda + \mu}{\lambda + 2
  \mu} \left[-\frac{\lambda + 2\mu}{\lambda + \mu} (e_{xx}^* -
  e_{yy}^*) \cos 2\theta - (e_{xx}^* + e_{yy}^*)\right]\\
\ad{e}_{\theta \theta} &=& \frac{S}{2 \pi r^2} \frac{\lambda + \mu}{\lambda + 2
  \mu} \left[ \frac{\lambda}{\lambda + \mu} (e_{xx}^* - e_{yy}^*) \cos
  2\theta + (e_{xx}^* + e_{yy}^*) \right]\\
\ad{e}_{r \theta} &=& \frac{S}{2 \pi r^2} \frac{\lambda + \mu}{\lambda + 2
  \mu} \left[-(e_{xx}^* -
  e_{yy}^*) \sin 2\theta \right]
\end{eqnarray*}

using $\ad{e}_{xx} - \ad{e}_{yy} = \cos (2 \theta) \left( \ad{e}_{r
  r} - \ad{e}_{\theta \theta} \right) - 2 \sin (2 \theta) \ad{e}_{r
  \theta}$, we obtain for the deviatoric strain:

\begin{equation}
\ad{e}_{xx} - \ad{e}_{yy} = \frac{S}{\pi r^2} \frac{\lambda +
  \mu}{\lambda + 2 \mu} \left[ - (e_{xx}^* - e_{yy}^*) \cos 4\theta -
  (e_{xx}^* + e_{yy}^*) \cos 2\theta\right] \label{eq:dev_2D}
\end{equation}

\section{Eshelby inclusion in an anisotropic material} \label{sec:Ap_aniso}
We consider the case of an orthorhombic bidimensional material which
free energy is of the form~\cite{Landau}:
\begin{equation*}
F = \frac{1}{2}\lambda_{xxxx} e_{xx}^2 + \frac{1}{2}\lambda_{yyyy}
e_{yy}^2 + \lambda_{xxyy} e_{xx} e_{yy} + 2 \lambda_{xyxy} e_{xy}^2
\end{equation*}
From the equilibrium equations $\frac{\partial \sigma_{ik}}{\partial
  x_k} + b_i = 0$, where $b_i$ is a body force, we can then deduce
\begin{eqnarray*}
\lambda_{xxxx} \frac{\partial^2 u_x}{\partial x^2} + \lambda_{xyxy}
\frac{\partial^2 u_x}{\partial y^2}  + (\lambda_{xxyy} +
\lambda_{xyxy}) \frac{\partial^2 u_y}{\partial x \partial y} + b_x &=&
0\\
\lambda_{yyyy} \frac{\partial^2 u_y}{\partial y^2} + \lambda_{xyxy}
\frac{\partial^2 u_y}{\partial x^2}  + (\lambda_{xxyy} +
\lambda_{xyxy}) \frac{\partial^2 u_x}{\partial x \partial y} + b_y &=&
0
\end{eqnarray*}
To find the Green's function of this system, i.e. the solution of the
system when $b_i = n_i \delta(\vec{r})$ where $\vec{n}$ is a vector of
unit norm, we work in the Fourier space. Noting the Fourier transform
of the components of the Green tensor $\mathbb{G}_{ij}$, we obtain the
following relations:
\begin{eqnarray*}
\lambda_{xxxx} q_x^2 \mathbb{G}_{xx} + \lambda_{xyxy} q_y^2
\mathbb{G}_{xx} + (\lambda_{xxyy} + \lambda_{xyxy}) q_x q_y
\mathbb{G}_{yx} &=& \frac{1}{4\pi^2}\\
\lambda_{xxxx} q_x^2 \mathbb{G}_{xy} + \lambda_{xyxy} q_y^2
\mathbb{G}_{xy} + (\lambda_{xxyy} + \lambda_{xyxy}) q_x q_y
\mathbb{G}_{yy} &=& 0\\
\lambda_{yyyy} q_y^2 \mathbb{G}_{yy} + \lambda_{xyxy} q_x^2
\mathbb{G}_{yy} + (\lambda_{xxyy} + \lambda_{xyxy}) q_x q_y
\mathbb{G}_{xy} &=& \frac{1}{4\pi^2}\\
\lambda_{yyyy} q_y^2 \mathbb{G}_{yx} + \lambda_{xyxy} q_x^2
\mathbb{G}_{yx} + (\lambda_{xxyy} + \lambda_{xyxy}) q_x q_y
\mathbb{G}_{xx} &=& 0
\end{eqnarray*}
The components $\mathbb{G}_{yx}$ can be deduced from those
relations. We compute them in the case of small anisotropy, i.e. when
$\lambda_{xxxx} - \lambda_{xxyy} - 2 \lambda_{xyxy}$ and
$\lambda_{yyyy} - \lambda_{xxyy} - 2 \lambda_{xyxy}$ are
small~\cite{Landau}. We then write:
$$\lambda_{xxxx} = \lambda + 2\mu - \epsilon \hspace{1cm}
\lambda_{yyyy} = \lambda + 2\mu + \epsilon \hspace{1cm}
\lambda_{xxyy} = \lambda \hspace{1cm} \lambda_{xyxy} = \mu$$
with $\epsilon \ll \lambda , \mu$. We obtain then
\begin{eqnarray*}
\mathbb{G}_{xx} &=& \frac{1}{4\pi^2 \mu} \left( \frac{1}{q^2} -
\frac{\lambda + \mu}{\lambda + 2\mu} \frac{q_x^2}{q^4}\right) \left[ 1
  + \frac{\epsilon}{\lambda + 2 \mu} \frac{q_x^2}{q^2} \left(1 +
  \frac{(\lambda + \mu) q_y^2}{(\lambda + \mu) q_y^2 + \mu q^2}
  \right) \right]\\
\mathbb{G}_{yy} &=& \frac{1}{4\pi^2 \mu} \left( \frac{1}{q^2} -
\frac{\lambda + \mu}{\lambda + 2\mu} \frac{q_y^2}{q^4}\right) \left[ 1
  - \frac{\epsilon}{\lambda + 2 \mu} \frac{q_y^2}{q^2} \left(1 +
  \frac{(\lambda + \mu) q_x^2}{(\lambda + \mu) q_x^2 + \mu q^2}
  \right) \right]\\
\mathbb{G}_{xy} = \mathbb{G}_{yx} &=& - \frac{1}{4\pi^2 \mu}
\frac{\lambda + \mu}{\lambda + 2\mu} \frac{q_x q_y}{q^4} \left[ 1 +
  \frac{\epsilon}{\lambda + 2 \mu} \frac{q_x^2 - q_y^2}{q^2} \right]
\end{eqnarray*}

We have $\hat{e}_{il}^* = 4 \pi^2 S q_k q_l \mathbb{G}_{ij}
\sigma_{jk}^*$~\cite{Picard2004} with $S$ the surface of the
inclusion, from which we deduce:
\begin{equation}
\hat{e}_{xx} - \hat{e}_{yy} = -\frac{S \sigma^*}{\mu} \left[ 1 -
  \frac{\lambda + \mu}{\lambda + 2\mu} \frac{(q_x^2 - q_y^2)^2}{q^4} +
  \frac{\epsilon}{\lambda + 2 \mu} \left( \frac{q_x^2 - q_y^2}{q^2} -
  \frac{\lambda + \mu}{\lambda + 2\mu} \frac{(q_x^2 - q_y^2)^3}{q^6}
  \right) \right]
\label{eq:aniso}
\end{equation}
for $\bm{\sigma^*} = \left(
\begin{array}{cc}
\sigma^* & 0\\
0 & -\sigma^*\\
\end{array}
\right)$.

\end{widetext}
The first isotropical term on the right hand side of
eq.~\ref{eq:aniso} Fourier transform leads to a Dirac term in $\vec{r}
= \vec{0}$. It is in fact due to the ponctual plastic event in the
inclusion and should be removed when computing the elastic
response~\cite{Picard2004}. The $\frac{(q_x^2 - q_y^2)^2}{q^4}$ leads
to a $\frac{\cos 4 \theta}{r^2}$ term in the real
space~\cite{Picard2004}. It differs from the usual form
$\frac{q_x^2q_y^2}{q^4}$ obtained in simple shear configuration
because the pattern is rotate of $45^{\circ}$ in a biaxial loading
compare to simple shear. We thus recover the quadrupolar response for
an isovolumic inclusion in absence of anisotropy ($\epsilon = 0$).

The term of order $\epsilon$ gives the modification of the response
due to the anisotropy of the matrix. From symetry arguments, we see
that those terms will give rise in the real space to terms of the form
$\frac{\cos 2 \theta}{r^2}$ and $\frac{\cos 6 \theta}{r^2}$.


\end{document}